# Ammonia Synthesis under Ambient Conditions: Insights into Water-Nitrogen-Magnetite Interfaces

Sruthy K. Chandy[1,2], Mauricio Lopez Luna[2], Nykita Z. Rustad[3], Isaac N. Zakaria[3,5], Andreas Siebert[2,3], Shane Devlin[3,6], Wan-Lu Li[7, 8], Monika Blum[2,3]*, Teresa Head-Gordon[1,2,4,5*]

[1]Kenneth S. Pitzer Theory Center and Department of Chemistry, University of California, Berkeley, CA, 94720 USA
[2]Chemical Sciences Division, Lawrence Berkeley National Laboratory, Berkeley, CA, 94720 USA
[3]Advanced Light Source, Lawrence Berkeley National Laboratory, Berkeley, CA, 94720 USA
[4]Departments of Bioengineering and [5]Chemical and Biomolecular Engineering, University of California, Berkeley, CA, 94720, USA
[6]Nevada Extreme Conditions Laboratory, University of Nevada, Las Vegas, NV, 89154, USA
[7]Aiiso Yufeng Li Family Department of Chemical and Nano Engineering and [8]Materials Science and Engineering, University of California, San Diego, La Jolla, CA 92093 USA

**ABSTRACT**
New routes for transforming nitrogen into ammonia at ambient conditions would be a milestone toward an energy efficient and economically attractive production route in comparison to the traditional Haber-Bosch process. Recently, the synthesis of ammonia from water and nitrogen at room temperature and atmospheric pressure has been reported to be catalyzed by $Fe_3O_4$ at the air-water interface. By integrating ambient pressure X-ray photoelectron spectroscopy and *ab initio* molecular dynamics and free energy calculations, we investigate the underlying thermodynamic mechanisms governing ammonia and hydrazine formation at the water-$Fe_3O_4$-nanoparticle interface. We find that, unlike pure $Fe_3O_4$ where $N_2$ can only interact with a limited number of Fe sites, hydroxylated species introduce large and diverse adsorption geometries where $N_2$ can bind through either Fe sites or Fe-OH groups, each of which are capable of independently facilitating proton-coupled electron transfer.

*corresponding author: thg@berkeley.edu, mblum@lbl.gov

TOC Graphic:

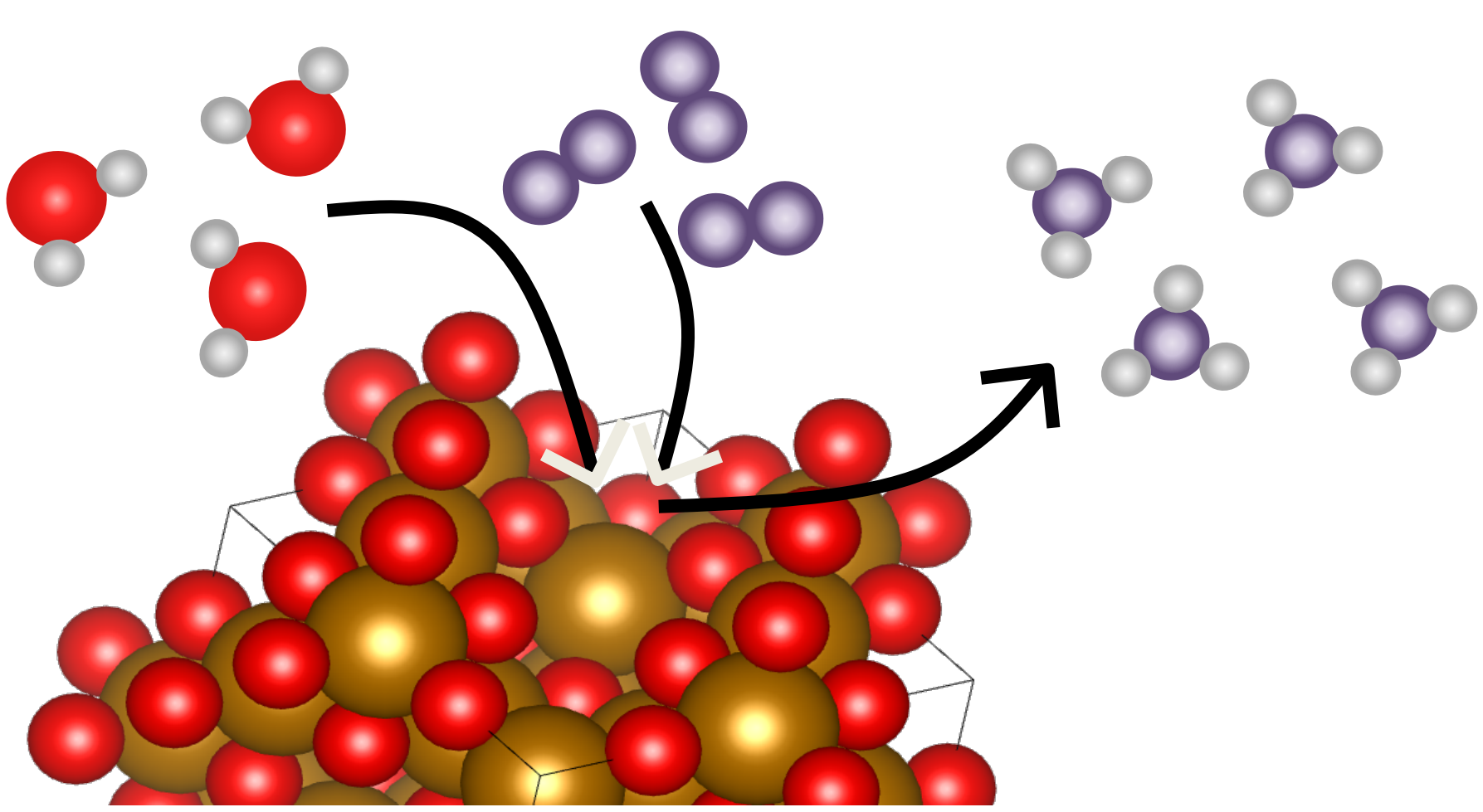

## INTRODUCTION

Ammonia ($NH_3$) is an essential compound with widespread applications in agriculture and chemical manufacturing[1], and has recently emerged as a promising hydrogen carrier due to its high energy density and favorable physicochemical properties that enables efficient transportation and storage.[2-4] Natural nitrogen ($N_2$) fixation primarily occurs under ambient conditions through microorganisms, which utilize Mo- and Fe-based nitrogenase enzymes to facilitate efficient $N_2$ reduction.[5-7] However, the synthetic production of ammonia is predominantly driven by the Haber-Bosch process ($N_2 + 3H_2 \rightarrow 2NH_3$), which operates under extreme conditions (~100 bar and ~700 K) and relies on high-purity hydrogen derived primarily from the steam reforming of fossil fuels. As a result, the Haber-Bosch process accounts for the emission of approximately 300 million metric tons of $CO_2$ emissions annually, underscoring its significant environmental footprint.[8] The discovery of new nitrogen fixation catalysts that operate at ambient conditions[9-13] could mitigate the challenges associated with the conventional Haber-Bosch process[8, 14] and provide a route to a more sustainable ammonia synthesis.

Inspired from the metal-complex centers of nitrogenase enzymes[15-16], ~30% of the electrocatalysts reported for the nitrogen reduction reaction (NRR) under ambient conditions are based on Fe and Mo.[5, 17-19] Recently the discovery of an alternative route for ammonia synthesis involving the combination of nitrogen gas with microdroplets of water with embedded magnetite ($Fe_3O_4$) nanoparticles has emerged as an intriguing new development.[20] This novel process operates at room temperature and atmospheric pressure without requiring electrochemical, thermal, or photochemical assistance, leveraging widely available natural resources such as water and air. Preliminary studies suggest that the redox properties of water microdroplet surfaces may play a pivotal role in facilitating the reaction.[21] Furthermore, experimental observations indicate the possible involvement of hydrazine as an intermediate or secondary product[20], though its precise role remains unclear. Remarkably, the reaction is robust even in the presence of compressed air[20], indicating that oxygen does not inhibit the process.

Despite these promising experimental findings, fundamental questions remain regarding the mechanistic role of $Fe_3O_4$ in nitrogen fixation at water interfaces under ambient conditions. Key uncertainties include the nature of active sites responsible for nitrogen activation, the influence of oxygen vacancies on catalytic efficiency, the role of water microdroplets in enabling the reaction, the presence and significance of reaction intermediates, particularly hydrazine. To address these challenges, this work aims to elucidate the interactions at the interface between water, nitrogen, and $Fe_3O_4$ nanoparticles. By integrating ambient pressure X-ray photoelectron spectroscopy (APXPS) and *ab initio* computational methods, we investigate the underlying reaction mechanisms governing ammonia and hydrazine formation.

## METHODS

**Ambient Pressure X-ray photoelectron spectroscopy.** Two sets of APXPS studies were performed at beamline 9.3.2 and 9.3.1 of the Advanced Light Source (ALS) at the Lawrence Berkeley National Laboratory and the experimental setups are described elsewhere.[22-24] For the 9.3.2 experiments the Fe 3p, N 1s, O 1s and C 1s were measured at a constant kinetic energy of 200 eV (hν = 255, 600, 730 and 485 eV respectively), which correspond to an approximately inelastic mean free path (IMFP) of 6.93 Å for $Fe_3O_4$. Additionally, the Fe 2p lines were collected at KE = 100 eV (hν = 810 eV, IMFP

≈ 5.41 Å) due to the monochromator limitations and excitation energy cut-off of 900 eV. The studied samples consisted of $Fe_3O_4$ nanoparticles ($Fe_3O_4$-NP; 50-100 nm particle size, Sigma Aldrich, 97%) drop-casted (suspended in milli-Q water) on a clean gold foil (0.5 mm thickness, Sigma Aldrich, 99.99%). The supported NPs were dried in air at room temperature for 1 hour before being introduced into the APXPS chamber.

The samples were studied in the as-prepared state in ultra-high vacuum (UHV) conditions as a reference. Then, the $Fe_3O_4$ nanoparticles were subject to a cleaning procedure ($5x10^{-7}$ Torr $O_2$ pressure and stepwise temperature increase from 298 to 623 K to remove adventitious carbon from the surface (**Supplementary Fig. S1a and S1b**) and the magnetite stoichiometry was closely monitored during the process.[25-26]. Once the sample was clean, a set of measurements was performed by exposing the samples to water vapor only (0.01 Torr, 0.1 Torr, and 0.3 Torr), and for each isobar experiment, the $Fe_3O_4$ nanoparticles were studied at different temperatures (between 284 and 345 K). In addition, the $Fe_3O_4$-NPs were exposed to mixtures of $N_2$:$H_2O$ (5:1, 1:2 and 1:6) at isobar conditions (0.06, 0.15 and 0.35 Torr respectively). During the whole study, additional mass spectroscopy data was collected in the second pumping stage of the electron analyzer (MKS e-vision 2).

In order to explore the effect of higher partial pressures of $N_2$ gas and $H_2O$ vapor on the $Fe_3O_4$-NPs, a second set of APXPS measurements was performed at the tender X-ray ambient pressure X-ray photoelectron spectroscopy beamline 9.3.1 (ALS, LBNL).[23] In these experiments, the samples were kept at room temperature and exposed to different $N_2$:$H_2O$ ratios; 1:1, 2:1, 5:1, 10:1, 1:5, 2:5, 1:1, and 2:1 (corresponding to 2, 3, 6, 11, 6, 7, 10 and 15 Torr respectively). A constant excitation energy of hν = 4000 eV was set to collect all photoelectron lines in this set of experiments. In both APXPS experiments, the main XPS analysis spot was chosen on an Au-free area and a secondary analysis spot containing mostly Au (and a very small fraction of $Fe_3O_4$-NP signal) was used for collecting the Au $4f_{7/2}$ peak position (83.93 eV) and used for beamline energy calibration and to exclude charging of the $Fe_3O_4$-NPs. Furthermore, the Fe 3p line was measured with every excitation energy for additional energy calibration. Supporting experimental information, including fitting procedures, is provided in the **Supplementary Information and Supplementary Figs. S1-S4**.

**Theoretical benchmarking for $Fe_3O_4$.** Accurately modeling magnetite NPs for *ab initio* molecular dynamics (AIMD) simulations is challenging due to the complex electronic structure of $Fe_3O_4$. Magnetite is characterized by its mixed-valence state, where $Fe^{3+}$ and $Fe^{2+}$ ions exist in a 2:1 ratio, contributing to its unique electronic and magnetic properties. To accurately describe the electronic structure and energetics of $Fe_3O_4$, highly correlated methods like full configuration interaction (CI)[27] or multireference approaches[28] are required. However, such methods are computationally prohibitive for AIMD simulations. Despite the limitations of density functional theory (DFT) in capturing strong electronic correlations, DFT provides a balance between computational cost and accuracy, making it a practical choice for to investigate nano-$Fe_3O_4$ catalysts for NRR mechanistic calculations.[29]

Magnetite exhibits multiple spin states corresponding to distinct multiplicities: M=15 (high-spin), M=7 (intermediate-spin), and M=5 (low-spin), arising from different spin arrangements of the Fe atoms (**Supplementary Fig. S6**). Identifying the correct ground spin state is critical, as it directly influences the stability and electronic properties of the system. To identify the appropriate spin state and functional for $Fe_3O_4$, we employed five DFT functionals spanning different rungs of "Jacob's Ladder" of approximations—PBE (GGA), BP86 (GGA), B97M-rV (meta-GGA), revPBE0-D3 (hybrid GGA with dispersion correction), and wB97X-V (range-separated hybrid) using Q-Chem[30].

Calculations were also performed with two basis sets, DZVP-MOLOPT and TZVP-MOLOPT, and the GTH-PBE pseudopotential using the CP2K[31] software package.

The calculated relative energies of the spin states M = 15, M = 7, and M = 5 for both basis sets are summarized in **Supplementary Table S1**. Key findings reveal that B97M-rV[32] and revPBE0-D3[33] consistently predict M = 15 (high-spin state) as the ground state across both basis sets, confirming the reliability of these functionals for $Fe_3O_4$. The BP86 functional[34] identifies M = 15 as the ground state when using the DZVP-MOLOPT basis set but slightly favors M = 7 by 0.02 eV with the TZVP-MOLOPT basis set. In contrast, PBE[35] predicts M = 5 (low-spin state) as the ground state, a result inconsistent with all other functionals, underscoring its limitations for strongly correlated systems like $Fe_3O_4$. While ωB97X-V[29] favors M = 7 as the lowest energy state for both basis sets, the energy difference relative to M = 15 is minimal (0.04 eV), falling within typical DFT error margins. Given the small energy difference (≤ 0.04 eV) between M = 15 and M = 7, the high-spin state M = 15 is justifiably chosen as the ground state for modeling the $Fe_3O_4$ system.

Thus to ensure robust and accurate results but affordable AIMD simulations, we selected the B97M-rV functional with rVV10 dispersion correction combined with the TZVP-MOLOPT basis set and GTH pseudopotential[36]. This choice is supported by the consistent prediction of M = 15 as the ground state across both basis sets and the demonstrated accuracy of B97M-rV in previous studies involving Fe systems, such as those observed in nitrogenase catalysis[37]. Unlike DFT+U methods, which are commonly applied to periodic bulk systems within the LDA or GGA framework to correct for issues like 3d electron delocalization, B97M-rV avoids the need for empirically tuned U parameters for describing electron delocalization in systems like $Fe_3O_4$.[38] Meta-GGA functionals like B97M-rV can often treat the electronic structure more accurately than LDA or GGA due to the inclusion of the Kohn-Sham kinetic energy density.

While the experimental nanoparticles likely have more complex surface environments (being larger, and having defect sites and oxygen vacancies) our use of the smallest representative unit of magnetite, $Fe_3O_4$, consisting of three multivalent iron atoms and four oxygen atoms, captures the essential mixed-valence and spin-state characteristics of the system while remaining computationally tractable. While we acknowledge that periodic slab models of $Fe_3O_4$ catalysts can represent extended surfaces or defect states, such systems are currently computationally prohibitive given that we extensively characterize their free energy pathways as described below.

**Ab Initio molecular dynamics (AIMD) and free energy calculation.** AIMD simulations were performed to study hydroxyl concentrations on $Fe_3O_4$ surfaces at four different temperatures of 288 K, 298 K, 313 K, and 333K using the CP2K[24] package. We used a time step of 0.5 fs in the NVT ensemble. Statistical average values were collected over 35 ps. The electron density was represented using the Gaussian and Plane Wave (GPW) method with a plane-wave cutoff of 400 Ry. The TZVP-MOLOPT basis set and GTH-PBE pseudopotentials were used for Fe, O, and H atoms, and the SCF convergence threshold was set to $10^{-6}$.

To ensure reliable treatment of boundary effects, we conducted a benchmark study to determine an appropriate simulation box size. We evaluated the $Fe_3O_4$ cluster with six water molecules in periodic boxes ranging from 8 Å to 25 Å (**Supplementary Fig. S7**). Geometry optimizations were performed using the B97M-rV functional with the TZVP-MOLOPT basis set and GTH-PBE pseudopotentials. Single-point energy calculations indicated that the energies converge at a box size of 10 Å, which was subsequently chosen for the AIMD simulations, ensuring that the observed structural dynamics are not artifacts of artificial confinement.

Simulations were initialized from well-relaxed B97M-rV optimized structures (spin multiplicity = 15), which ensured accurate bonding geometries prior to the onset of dynamics. The electronic structure was treated using the meta-GGA B97M-rV functional combined with the rVV10 nonlocal dispersion correction, which has been shown to reliably model redox-active transition metal and metalloenzyme systems.[39-40]

To assess the onset and persistence of surface hydroxylation during AIMD, we tracked the time evolution of Fe–OH and Fe–O bonding motifs identified using distance cutoffs derived from optimized B97M-rV/TZVP-MOLOPT geometries of hydroxylated $Fe_3O_4$ species: 0.96 Å for the O–H bond and 1.96 Å for the Fe–O bond. To account for thermal fluctuations, we applied a ±0.05 Å buffer, resulting in thresholds of ≤1.1 Å for O–H and ≤2.1 Å for Fe–O. The time evolution of Fe–OH formation, quantified using this criterion, strongly supports our proposed mechanism in which hydroxylation precedes and facilitates nitrogen activation.

To further evaluate the observed trends in the XPS experiments with magnetite and water, we calculated the free energy changes for water dissociation events leading to the formation of hydroxyl species on the $Fe_3O_4$ surface. Free energy calculations were performed using Q-Chem, employing the B97M-rV functional with the def2-TZVPP basis set. All possible proton coupled electron addition steps for $Fe_3O_4$ with the presence of water are explored, and enthalpies and free energies are calculated for all reaction steps.

## RESULTS

**Figure 1** shows the Fe 2p (a) and O 1s (b) APXPS spectra of the $Fe_3O_4$-NP sample. The sample was investigated in an as-prepared state under UHV condition and then underwent a cleaning process (described in methods) before being exposed to a constant water vapor pressure (here shown at 0.3 Torr) with varying temperatures from 284, 298, 318, to 345 K. The Fe 2p spectra (**Figure 1(a) and Supplementary Fig. S1**) reveal the classical chemical composition of magnetite. i.e., $Fe^{2+}(Fe^{3+})_2(O^{2-})_4$, containing both ferrous and ferric iron.[41] During the whole experiment the peak positions and $Fe^{3+}$ : $Fe^{2+}$ ratios stayed constant (1.95 ± 0.9) and the spectra were fitted with a $Fe^{3+}$ and $Fe^{2+}$ doublet as well as the corresponding satellites (**see Supplementary Information and Figure 1(a) bottom**) only, leading to the conclusion that the $Fe_3O_4$-NP surface is not influenced by the surrounding experimental conditions. In addition, a careful beam damage study was performed in the beginning of the experiments as well as a control measurement at the end to exclude chemical changes due to the synchrotron radiation.

The O 1s APXPS spectra of the same $Fe_3O_4$-NP sample are shown in **Figure 1(b)**. The as-prepared sample in UHV at RT conditions (bottom), has a shoulder to higher binding energies of the Fe-O peak at 529.9 eV, which can be assigned to surface adsorbates due to the drop casting process. In this case there is a C-O component at 531.2 eV as well as a potential Fe-OH component at 531.7 eV (**Supplementary Fig. S2(a)**). The surface adsorbates can be removed after gradually heating the sample to 623 K (**Figure 1(b) and Supplementary Fig. S2 (b)**) and exposing it to a partial $O_2$ gas pressure of $5x10^{-7}$ Torr. Although, after cooling back down to RT, a small carbon contribution is reappearing due to the nature of the NPs and is accounted for in the O 1s fits throughout the experiments **(Supplementary Fig. S2)**.

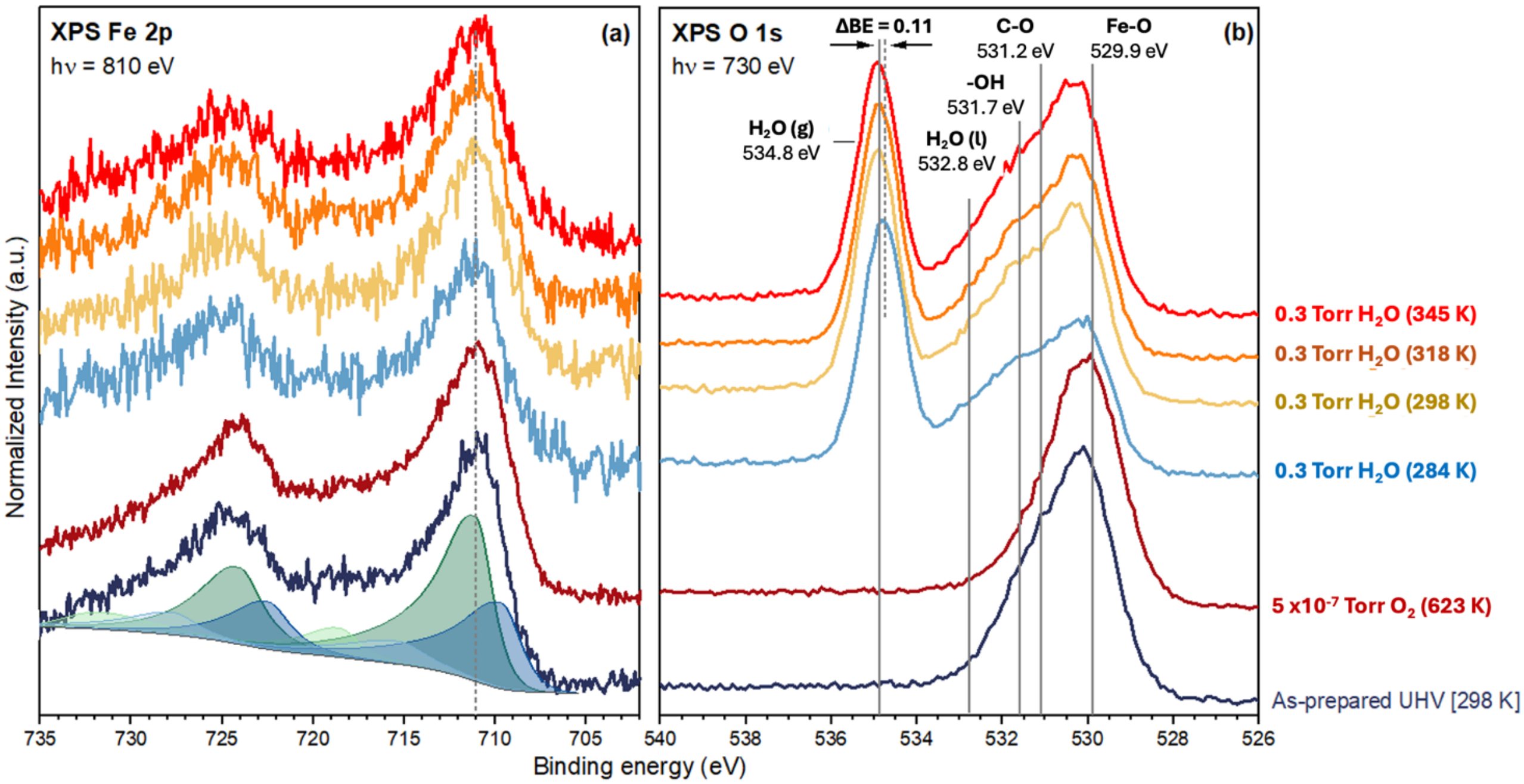

**Figure 1**. *APXPS core level spectra of the $Fe_3O_4$ nanoparticles when exposed to water as a function of temperature.* **(a)** APXPS Fe 2p core level spectra of the $Fe_3O_4$ nanoparticles (KE = ~100 eV). A fit was added for the as-prepared UHV spectrum **(b)** corresponding O 1s spectra of the $Fe_3O_4$ nanoparticles at the same conditions (KE = ~200 eV).

When exposing the clean $Fe_3O_4$-NPs to 0.3 Torr of water vapor three species emerge in the O 1s spectrum in addition to the lattice O 1s peak of the NPs, as illustrated in **Figure 1(b)**, with peak positions at 534.8 eV, 532.8 eV, and 531.7 eV. The first peak corresponds to water vapor in the chamber, which gets excited by the synchrotron radiation between the sample surface and the analyzer nozzle and is not pinned to the Fermi level and thus can have slight shifts in binding energy. The peak at 532.8 eV is associated with surface-bond liquid water, and the third peak can be attributed to excess hydroxyl species on the sample surface. The findings are in excellent agreement with previous water vapor interaction studies on a $Fe_3O_4$ (001) single crystal.[37]

The peak areas of the Fe 2p and O 1s spectra (**Supplementary Fig. S1 and S2**) were monitored under similar conditions and the interactions between the $Fe_3O_4$-NPs and water vapor were investigated at increasing temperatures. **Figure 2(a)** shows the component area ratios of $OH^-$/Fe-O (closed triangles) and $H_2O$(l)/Fe-O (open triangles; corresponding to the hydroxyl groups as well as adsorbed liquid water on the NP surface, respectively) derived from the O 1s spectra measured with 5 K temperature steps from 284 to 345 K and at 0.3 Torr constant vapor pressure. The spectra show an increasing $OH^-$/Fe-O area ratio with temperature indicating a higher concentration of hydroxyl groups forming on the magnetite surface, with an unexpected and, to the authors knowledge, unreported in literature “jump” around room temperature (RT). Conversely, the relative ratio of $H_2O$(l)/Fe-O decreases with temperature, showing a reduction in liquid water concentration at increasing temperatures, which is a previous reported trend for water adsorption studies on metal oxides.[37, 42]

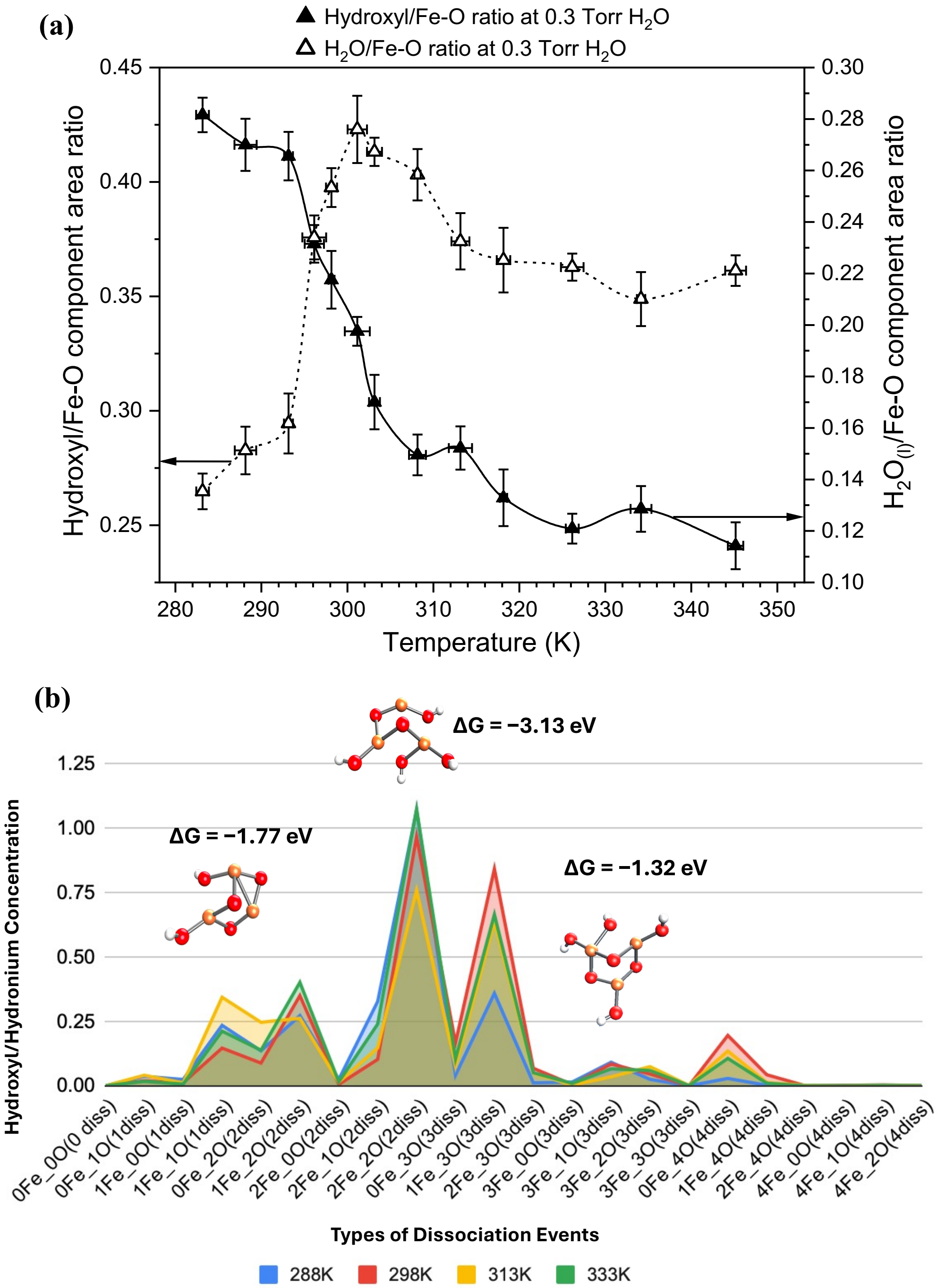

**Figure 2**. *APXPS core level spectra and ab initio molecular dynamics of the $Fe_3O_4$ nanoparticles when exposed to water as a function of temperature.* (a) Experimental component area ratios for $OH^-$:Fe-O (closed symbols) and $H_2O$(l):Fe-O (open symbols) as a function of temperature. Peak fitting details for Fe 2p and O 1s can be found in Supplementary Information. (b) AIMD trends with temperature to monitor the concentration of water dissociation events on the magnetite nanoparticle. The types of dissociation events are represented as xFe–yO, where *x* denotes the number of OH fragments attached to Fe sites, and *y* indicates the number of H atoms bound to surface O atoms of the nanoparticle. The values in parentheses specify the number of water dissociation events. The criteria for identifying these species are determined using a cutoff distance of ±0.05 Å from the optimized structure's bond lengths. Example free energies evaluated for different water splitting events.

Additionally, during the isobar series it was observed that higher vapor pressures are associated with larger $OH^{-}$/Fe-O component area ratios, i.e., the vapor pressure further promotes the formation of hydroxyl groups on the surface (**Supplementary Fig. S3a**). When the $Fe_3O_4$-NPs sample is introduced to water vapor and an additional 0.05 Torr of $N_2$ gas, the same dissociation trend for the $H_2O(l)$/Fe-O component area ratio can be observed as for water vapor only measurements but the $OH^{-}$/Fe-O ratio shows a constant increase with increasing temperature and no indication of a "jump" at RT, especially visible at the 0.3 Torr $H_2O$ + 0.05 Torr $N_2$ isobar (**Supplementary Fig. S3b**). In addition, the constant $Fe^{3+}$:$Fe^{2+}$ ratio of the Fe 2p spectra ($H_2O$+$N_2$; not shown) indicates that the oxidation state of the $Fe_3O_4$-NPs is not influenced by the addition of $N_2$ gas to the experimental conditions.

To understand these trends, we employed a cluster model of $Fe_3O_4$ with six water molecules to mimic the low water content of the experiments and performed AIMD simulations in the NVT ensemble at the same four temperatures of the experiment (see Methods for further details). As seen in **Figure 2b**, we find that the magnetite nanoparticle is quite active and actively promotes water dissociation to bound hydroxide and hydronium ions to magnetite, with enthalpy changes showing that water dissociation on the iron oxide surface is exothermic. Notably, the hydrogen atoms bound to oxygen on $Fe_3O_4$ and the hydroxyl groups attached to iron sites of $Fe_3O_4$ are indistinguishable when analyzed using XPS spectroscopy, as their energy signatures strongly overlap. Overall, the temperature trends are explained quite easily by the fact that all temperatures exhibit at least 2 complete water dissociation events, whereas only at temperatures above 288K we see significant numbers of 3 and even 4 water dissociations, leading to a greater concentration of surface hydroxyl groups relative to intact Fe–O bonds as seen experimentally in **Figure 2b**.

To assess the onset and persistence of surface hydroxylation during AIMD, we tracked the time evolution of Fe–OH and Fe–O bonding motifs over a 35 ps trajectory at 298 K (**Supplementary Fig. S8**). We observed that hydroxylation begins spontaneously within the first 2–5 ps and remains stable throughout the trajectory. These events coincide with a decrease in intact Fe–O bonds, indicating irreversible bond cleavage and structural reorganization at the surface. No reversals of dissociation were observed within this time frame, suggesting dynamic stabilization of the hydroxylated states. These dynamic signatures serve as direct mechanistic indicators of interfacial activity and support the role of hydroxylation as a precursor to nitrogen activation.

Of great interest to this study is the N 1s core electron spectrum region and a possible $NH_3$ peak signature. Due to the low signal - to - noise ratio of the soft X-ray experiments in the utilized vapor/gas pressure regime, the N 1s spectra were recorded with tender X-ray APXPS allowing higher gas pressures (**Figure 3 and Supplementary Fig. S4**). In the presence of $N_2$ gas and water vapor the N 1s spectra of the $Fe_3O_4$-NPs sample show a very intensive peak at 405 eV, which can be associated with free $N_2$ gas in the chamber between the sample surface and the electron analyzer (similar to the water gas peak in the O 1s spectra) excited by the X-rays. Three peaks emerge with building pressures at 401.1, 399.9, and 399.0 eV. The latter is expected to be a $Fe_0^{3+}$-N component, and its area doesn't change over the different pressure regimes, indicating a saturation at already 1 Torr of $N_2$ gas pressure. The peak has previously been reported in a low pressure $N_2$ adsorption study on a $Fe_3O_4$ (001) single crystal, where it was promoted by a $H_2$ reduction of the single crystal.[43] The same study also reported an N-O component at 400.6 eV, which in this study couldn't be clearly identified. In the literature this peak is sometimes identified as adsorbed $N_2$ gas.[44] The peaks at 401.1 and 399.9 eV can be associated with hydrazine and ammonia, respectively. The interaction of hydrazine with different metals has been well studied in literature as well as the adsorption and decomposition thereof.[45-48] These findings are in excellent agreement with our peak position. **Supplementary Figure S5** shows additional mass spectroscopy data from the soft X-ray experiments. Both masses of interest for ammonia (m/z=17) and hydrazine (m/z=32) overlap with the masses of $OH^-$ and $O_2$, respectively. For m/z 17 there is no clear trend visible due to the change in $OH^-$ formation with different experimental conditions (**Figure 2a** and **Supplementary Fig. S3**), in contrast, m/z=32 shows a clear increase in pressure when adding $N_2$ gas to the $H_2O$ vapor, indicating the formation of hydrazine.

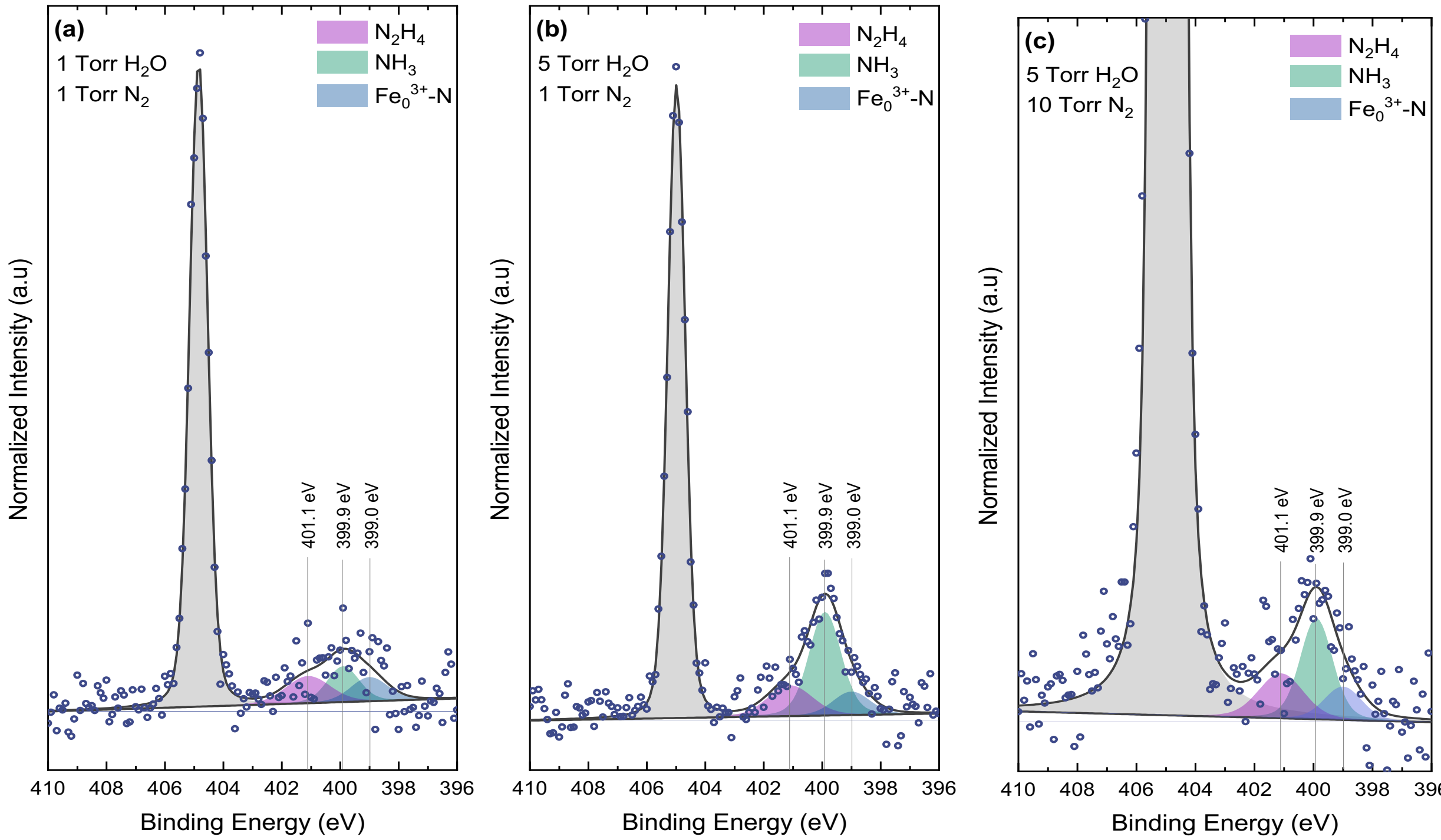

**Figure 3:** *Normalized N 1s core spectra with changes in gas pressures*. (a) 1 Torr $H_2O$ + 1 Torr $N_2$, (b) 5 Torr $H_2O$ + 1 Torr $N_2$, and (c) 5 Torr $H_2O$ + 10 Torr $N_2$.

In **Figure 3a**, the N 1s spectra were measured with equal pressures of $N_2$ gas and water vapor (1 Torr each), while in **Figure 3b** the water concentration was increased fivefold, keeping the $N_2$ concentration constant. Interestingly, increasing the water concentration led to a noticeable increase in the signatures of the ammonia species but not in the hydrazine component. To explore the effect of nitrogen concentration on the product peaks, a third experiment was conducted with the water vapor pressure being held constant at 5 Torr, and the $N_2$ gas pressure was increased to 10 Torr (**Figure 3c**). Here, surprisingly, no significant change in the ammonia signature was observed and only a small increase in the hydrazine signature compared to **Figure 3b**, where the $N_2$ concentration was 1 Torr. This finding demonstrates that the water concentration directly impacts the formation of ammonia and hydrazine, while the nitrogen concentration does not significantly influence the product peaks. The above results suggest that water is the limiting factor for the formation of ammonia and hydrazine on magnetite nanoparticles.

Using free energy calculations, we find that the energy gain of $N_2$ attachment to the Fe site of the bare $Fe_3O_4$ is calculated to be –0.46 eV, which is significantly less favorable than the energy gain associated with water dissociation and the absorption of hydroxyl and hydronium to the iron oxide nanoparticle as given in **Figure 2**. Furthermore, adding a second $N_2$ is thermodynamically uphill for pure $Fe_3O_4$ or $Fe_3O_4$ after one and two water dissociations with a free energy change of –0.37, 0.08, and 0.25 eV compared with one $N_2$ addition with a free energy change of –0.46, 0.005, and 0.02 eV (**Supplementary Fig. S6**). This is in line with the experiment, adding excess $N_2$ doesn't influence the concentration of ammonia or hydrazine, as it is thermodynamically unfavorable compared to single $N_2$ additions. This supports that the first step in the NRR on $Fe_3O_4$ surfaces is the formation of a water-dissociated species, rather than direct $N_2$ attachment to the bare iron oxide nanoparticle, which is consistent with the dependence on water content seen experimentally.

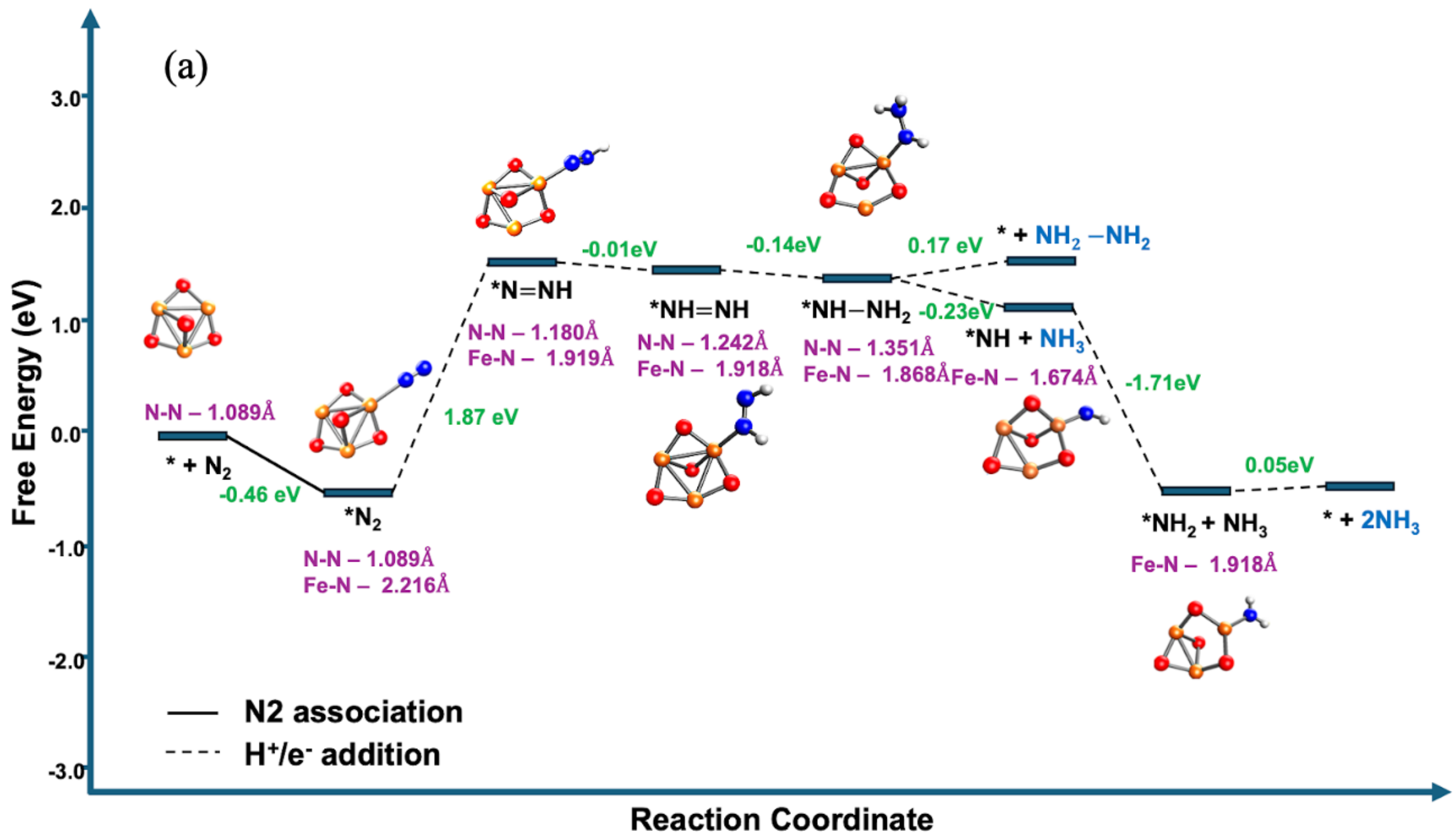

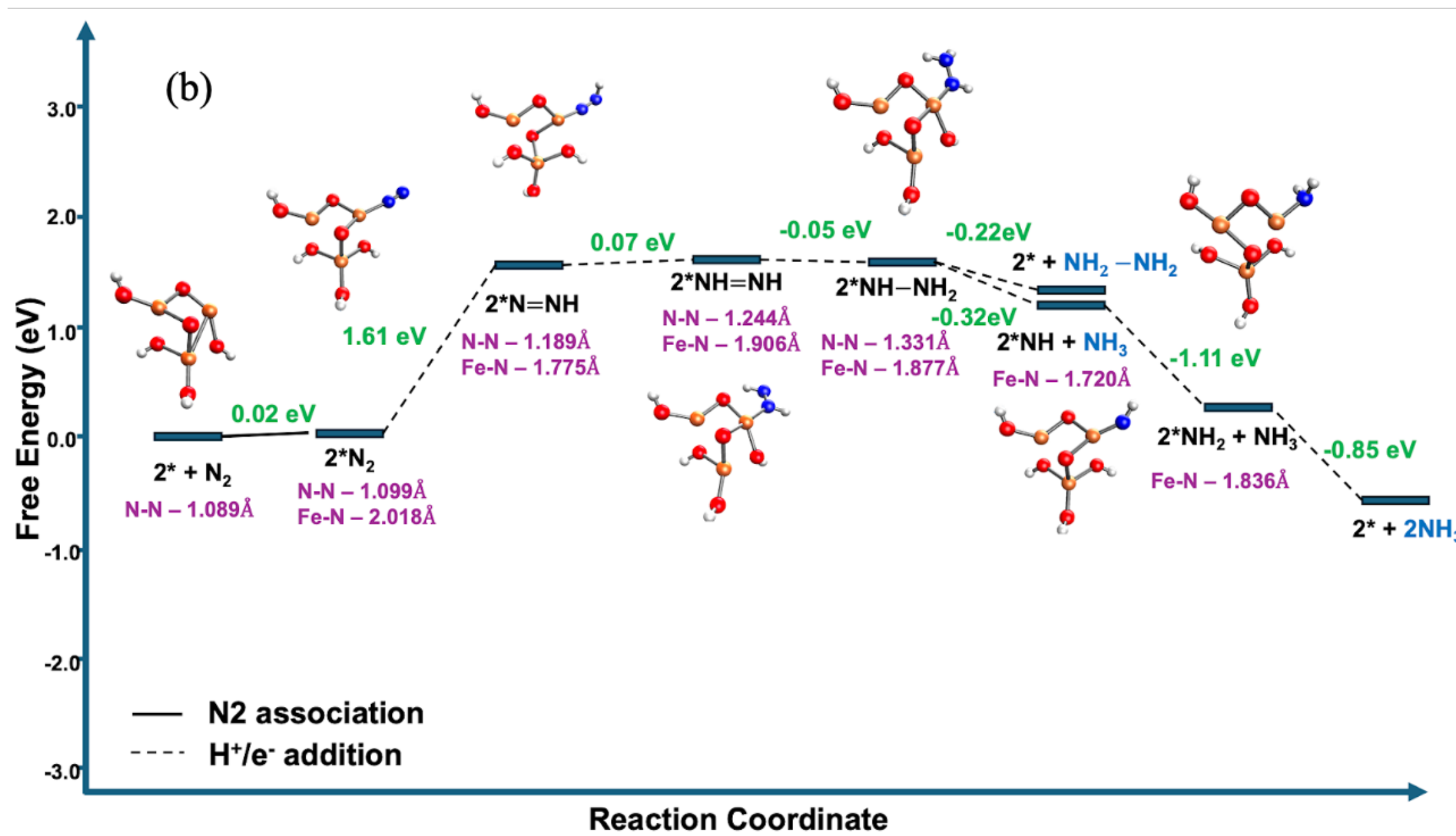

**Figure 4:** *Free energy change for the NRR mechanism on iron oxide with and without water.* (a) the bare $Fe_3O_4$ (*) and (b) iron oxide after two water dissociation (2* refers to $Fe_3(OH)_2O_4(H)_2$).

We assume that at the water-magnetite interface, a proton-coupled electron transfer (PCET) mechanism plays a central role, with nanoscale interfacial environments fostering high concentrations of protons ($H_3O^+$) and solvated electrons ($e^-$). With these assumptions, we examine two free energy pathways for the NRR in **Figure 4** for the bare iron oxide nanoparticle compared to the most observed two water dissociation species as seen in the AIMD simulations in **Figure 2**. The corresponding enthalpy plots are provided in **Supplementary Figs. S7 and S8**. We see that hydroxylation lowers the reaction free energy barriers by close to 0.3 eV for the rate determining formation of the N=NH species that activates the nitrogen bond. Similarly, the formation of the key $NH_2$-$NH_2$ intermediate is an uphill process on the bare $Fe_3O_4$ surface ($\Delta G$ = 0.17 eV) whereas it is a downhill step on the $Fe_3(OH)_2O_4(H)_2$ surface ($\Delta G$ = −0.22 eV), allowing hydrazine formation to proceed with minimal energetic resistance. Furthermore, the first $NH_3$ release is comparable between the two systems but is highly exergonic for the final ammonia product for the hydroxylated system ($\Delta G$ = −0.85 eV), making the final desorption step far more favorable than in the pure $Fe_3O_4$ state ($\Delta G$ = +0.05 eV for *). Similar conclusions are drawn from additional hydroxylation pathways as seen in **Supplementary Figs. S10-S17**.

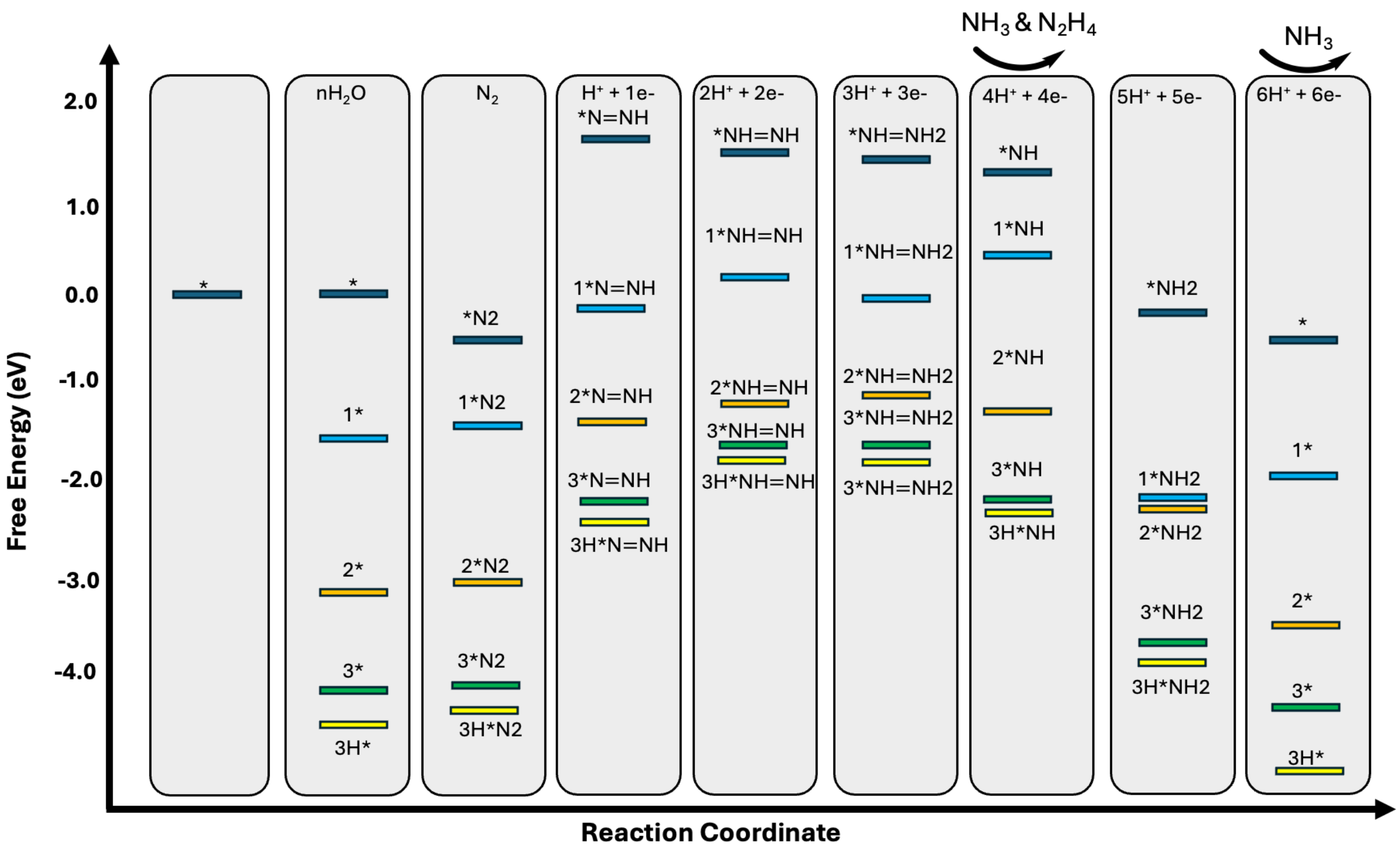

**Figure 5.** Energetics for the NRR pathway on bare $Fe_3O_4$ (*) and hydroxylated $Fe_3O_4$ species (1*, 2*, 3*, and 3H*). $nH_2O$ represents the number of water molecules dissociated on the *, where n goes from 0 to 3 to form *, 1*, 2*, 3*and 3H*. Detailed pathways of each species is given in **Figure 4** and **Supplementary Figs:S7-S10**

These results demonstrate that the hydroxylated species give rise to more favorable hydrazine and ammonia formation steps in agreement with the APXPS results that showed that water, not nitrogen, increased the NRR products. Furthermore, unlike pure $Fe_3O_4$ where $N_2$ can only interact with a limited number of Fe sites, hydroxylated species introduce large and diverse adsorption geometries where $N_2$ can bind through either Fe sites or Fe-OH groups, each of which are capable of independently facilitating proton-coupled electron transfer as showed in combined pathway as in **Figure 5**. This diversity prevents a single rate-determining step from dominating the reaction mechanism, and results in a more distributed and efficient hydrogenation process, leading to a faster conversion of $N_2$ into $NH_3$ and $N_2H_4$.

## DISCUSSION AND CONCLUSIONS

This combined experimental and theoretical study provides insight into the $Fe_3O_4$–$H_2O$–$N_2$ interface and its catalytic potential for ammonia and hydrazine synthesis. It corroborates previous studies showing ammonia synthesis from sprayed water microdroplets on an iron oxide catalyst and a Nafion-coated graphite mesh. This previous study utilizing mass spectroscopy proposed that the abundance of $H_3O^+$ and $e^-$ at the microdroplet interface facilitates the hydrogenation of $N_2$ bound to the $Fe_3O_4$ surface.[20] Here we have performed APXPS experiments under precisely controlled temperature and pressure conditions, which are further supported by ab initio calculations that provide mechanistic insights into the NRR. Specifically, the N 1s spectral signatures corresponding to ammonia and hydrazine show a dependence regarding the water vapor pressure but are independent from the $N_2$ gas pressure.

Ab initio molecular dynamics and free energy calculations found that water dissociation and the formation of hydroxyl species on iron oxide surfaces is preferred over nitrogen absorption. Additionally, key intermediates along the NRR pathways are far more stable in the hydroxylated form versus the bare nanoparticle. The formation of hydroxylated species leads to a higher density of Fe-N reaction centers, effectively increasing the probability of $N_2$ interaction and reduction. The enhanced hydrogenation efficiency collectively results in greater production of ammonia and hydrazine, which explains the observed increase in nitrogen-containing species in the XPS spectra. This agreement between theoretical predictions and experimental observations confirms that hydroxylation plays a crucial role in tuning $Fe_3O_4$-NPs catalytic activity even at room temperature, enabling more nitrogen reduction pathways as water content increases. At the water-magnetite interface, a PCET mechanism plays a central role, with nanoscale interfacial environments fostering high concentrations of protons ($H_3O^+$) and solvated electrons ($e^-$), much like that in the nitrogenase enzyme[15].

These surface-mediated processes on $Fe_3O_4$-NPs highlight the unique catalytic environment created by the interplay of localized potentials, active sites, and interfacial dynamics, laying the fundamental groundwork for future advancements in environmentally benign ammonia synthesis. But at present there is no experimental kinetic data for $Fe_3O_4$-catalyzed ammonia formation under these ambient conditions, likely because product yields are low and exquisite detection sensitivity is required. Instead this work has clarified the thermodynamic feasibility of a large number of available pathways, whose diversity creates a reaction mechanism that deserves a better kinetic understanding. We view this work as a foundational step toward a more complete kinetic description, which we hope to actively pursue in follow-up studies.

**ACKNOWLEDGMENTS**

We thank the CPIMS program, Office of Basic Energy Sciences, Chemical Sciences Division of the U.S. Department of Energy under Contract DE-AC02-05CH11231 for funding of the experimental and theoretical studies described here. N. I. Z. was supported in part by an Advanced Light Source (ALS) Doctoral Fellowship in Residence and a National Defense Science and Engineering Graduate (NDSEG) Fellowship, for which we thank the U.S. Department of Energy Office of Science and the U.S. Department of Defense, respectively. The APXPS measurements used the Advanced Light Source, which is a U.S. Department of Energy Scientific User Facility under contract no. DE-AC02-05CH11231. This work used computational resources provided by the National Energy Research Scientific Computing Center (NERSC), a U.S. Department of Energy Office of Science User Facility operated under Contract DE-AC02-05CH11231, and the Lawrencium computational cluster resource provided by the IT Division at the Lawrence Berkeley National Laboratory (Supported by the Director, Office of Science, Office of Basic Energy Sciences, of the U.S. Department of Energy under Contract No. DE-AC02-05CH11231).

**SUPPORTING INFORMATION**

Further description of experimental and theoretical methods, peak fitting details, mass spectroscopy data and additional APXPS core level spectra, benchmarks for DFT, spin state analysis, and additional free energy and enthalpy pathways for NRR with iron oxide.

**REFERENCES**

1. Pattabathula, V.; Richardson, J., Introduction to Ammonia Production. *Chem. Eng. Prog* **2016,** *112* (9), 69–75-69–75.
2. Christensen, C. H.; Johannessen, T.; Sørensen, R. Z.; Nørskov, J. K., Towards an Ammonia-Mediated Hydrogen Economy? Catal. *Today* **2006,** *111* (1), 140–144-140–144.
3. Kong, J.; Lim, A.; Yoon, C.; Jang, J. H.; Ham, H. C.; Han, J.; Nam, S.; Kim, D.; Sung, Y. E.; Choi, J.; Park, H. S., Electrochemical Synthesis of NH3 at Low Temperature and Atmospheric Pressure Using a γ-Fe2O3 Catalyst. *ACS Sustain. Chem. Eng* **2017,** *5* (11), 10986–10995-10986–10995.
4. Morlanés, N.; Katikaneni, S. P.; Paglieri, S. N.; Harale, A.; Solami, B.; Sarathy, S. M.; Gascon, J., A Technological Roadmap to the Ammonia Energy Economy: Current State and Missing Technologies. *Chem. Eng. J* **2021,** *408*, 127310-127310.
5. Chen, J.; Cheng, H.; Ding, L. X.; Wang, H., Competing Hydrogen Evolution Reaction: A Challenge in Electrocatalytic Nitrogen Fixation. *Mater. Chem. Front* **2021,** *5* (16), 5954–5969-5954–5969.
6. Li, H.; Gu, S.; Sun, Z.; Guo, F.; Xie, Y.; Tao, B.; He, X.; Zhang, W.; Chang, H., The In-Built Bionic “MoFe Cofactor” in Fe-Doped Two-Dimensional MoTe2 Nanosheets for Boosting the Photocatalytic Nitrogen Reduction Performance. *J. Mater. Chem. A* **2020,** *8* (26), 13038–13048-13038–13048.
7. Liu, A.; Yang, Y.; Ren, X.; Zhao, Q.; Gao, M.; Guan, W.; Meng, F.; Gao, L.; Yang, Q.; Liang, X.; Ma, T., Current Progress of Electrocatalysts for Ammonia Synthesis Through Electrochemical Nitrogen Reduction Under Ambient Conditions. *ChemSusChem* **2020,** *13* (15), 3766–3788-3766–3788.

8. Cherkasov, N.; Ibhadon, A. O.; Fitzpatrick, P., A Review of the Existing and Alternative Methods for Greener Nitrogen Fixation. *Chem. Eng. Process. Process Intensif* **2015,** *90*, 24–33-24–33.
9. Ali, M.; Zhou, F.; Chen, K.; Kotzur, C.; Xiao, C.; Bourgeois, L.; Zhang, X.; MacFarlane, D. R., Nanostructured Photoelectrochemical Solar Cell for Nitrogen Reduction Using Plasmon-Enhanced Black Silicon. *Nat. Commun* **2016,** *7* (1), 11335-11335.
10. Li, M.; Huang, H.; Low, J.; Gao, C.; Long, R.; Xiong, Y., Recent Progress on Electrocatalyst and Photocatalyst Design for Nitrogen Reduction. *Small Methods* **2019,** *3* (6), 1800388-1800388.
11. Liu, X.; Jiao, Y.; Zheng, Y.; Jaroniec, M.; Qiao, S. Z., Building Up a Picture of the Electrocatalytic Nitrogen Reduction Activity of Transition Metal Single-Atom Catalysts. *J. Am. Chem. Soc* **2019,** *141* (24), 9664–9672-9664–9672.
12. Qiu, W.; Xie, X. Y.; Qiu, J.; Fang, W. H.; Liang, R.; Ren, X.; Ji, X.; Cui, G.; Asiri, A. M.; Cui, G.; Tang, B.; Sun, X., High-Performance Artificial Nitrogen Fixation at Ambient Conditions Using a Metal-Free Electrocatalyst. *Nat. Commun* **2018,** *9* (1), 3485-3485.
13. Xiao, L.; Zhu, S.; Liang, Y.; Li, Z.; Wu, S.; Luo, S.; Chang, C.; Cui, Z., Effects of Hydrophobic Layer on Selective Electrochemical Nitrogen Fixation of Self-Supporting Nanoporous Mo4P3 Catalyst under Ambient Conditions. *Appl. Catal. B Environ* **2021,** *286*, 119895-119895.
14. Humphreys, J.; Lan, R.; Tao, S., Development and Recent Progress on Ammonia Synthesis Catalysts for Haber–Bosch Process. *Adv. Energy Sustain. Res* **2021,** *2* (1), 2000043-2000043.
15. Li, W.-L.; Li, Y.; Li, J.; Head-Gordon, T., How thermal fluctuations influence the function of the FeMo cofactor in nitrogenase enzymes. *Chem Catalysis* **2023,** *3* (7), 100662.
16. Seefeldt, L. C.; Yang, Z.-Y.; Lukoyanov, D. A.; Harris, D. F.; Dean, D. R.; Raugei, S.; Hoffman, B. M., Reduction of Substrates by Nitrogenases. *Chemical Reviews* **2020,** *120* (12), 5082-5106.
17. Suryanto, B. H. R.; Du, H. L.; Wang, D.; Chen, J.; Simonov, A. N.; MacFarlane, D. R., Challenges and Prospects in the Catalysis of Electroreduction of Nitrogen to Ammonia. *Nat. Catal* **2019,** *2* (4), 290–296-290–296.
18. Wang, F.; Xia, L.; Li, X.; Yang, W.; Zhao, Y.; Mao, J., Nano-Ferric Oxide Embedded in Graphene Oxide: High-Performance Electrocatalyst for Nitrogen Reduction at Ambient Condition. *ENERGY Environ. Mater* **2021,** *4* (1), 88–94-88–94.
19. Zhou, F.; Azofra, L. M.; Ali, M.; Kar, M.; Simonov, A. N.; McDonnell-Worth, C.; Sun, C.; Zhang, X.; MacFarlane, D. R., Electro-Synthesis of Ammonia from Nitrogen at Ambient Temperature and Pressure in Ionic Liquids. *Energy Environ. Sci* **2017,** *10* (12), 2516–2520-2516–2520.
20. Song, X.; Basheer, C.; Zare, R. N., Making Ammonia from Nitrogen and Water Microdroplets. *Proc. Natl. Acad. Sci* **2023,** *120* (16), 2301206120-2301206120.
21. LaCour, R. A.; Heindel, J. P.; Zhao, R.; Head-Gordon, T., The Role of Interfaces and Charge for Chemical Reactivity in Microdroplets. *J Am Chem Soc* **2025,** *147* (8), 6299-6317.
22. Grass, M. E.; Karlsson, P. G.; Aksoy, F.; Lundqvist, M.; Wannberg, B.; Mun, B. S.; Hussain, Z.; Liu, Z., New ambient pressure photoemission endstation at Advanced Light Source beamline 9.3.2. *Review of Scientific Instruments* **2010,** *81* (5), 053106.
23. Axnanda, S.; Crumlin, E. J.; Mao, B.; Rani, S.; Chang, R.; Karlsson, P. G.; Edwards, M. O. M.; Lundqvist, M.; Moberg, R.; Ross, P.; Hussain, Z.; Liu, Z. Using "Tender" X-ray Ambient Pressure X-Ray Photoelectron Spectroscopy as A Direct Probe of Solid-Liquid Interface *Sci Rep* [Online], 2015, p. 9788.
24. Crumlin, E. J.; Liu, Z.; Bluhm, H.; Yang, W.; Guo, J.; Hussain, Z., X-ray spectroscopy of energy materials under in situ/operando conditions. *Journal of Electron Spectroscopy and Related Phenomena* **2015,** *200*, 264-273.

25. Kaya, S.; Ogasawara, H.; Nilsson, A., Determination of the surface electronic structure of Fe3O4(111) by soft X-ray spectroscopy. *Catalysis Today* **2015,** *240*, 184-189.
26. Kraushofer, F.; Mirabella, F.; Xu, J.; Pavelec, J.; Balajka, J.; Müllner, M.; Resch, N.; Jakub, Z.; Hulva, J.; Meier, M.; Schmid, M.; Diebold, U.; Parkinson, G. S., Self-limited growth of an oxyhydroxide phase at the Fe3O4(001) surface in liquid and ambient pressure water. *The Journal of Chemical Physics* **2019,** *151* (15), 154702.
27. David Sherrill, C.; Schaefer, H. F., The Configuration Interaction Method: Advances in Highly Correlated Approaches. In *Advances in Quantum Chemistry*, Löwdin, P.-O.; Sabin, J. R.; Zerner, M. C.; Brändas, E., Eds. Academic Press: 1999; Vol. 34, pp 143-269.
28. Vitillo, J. G.; Cramer, C. J.; Gagliardi, L., Multireference Methods are Realistic and Useful Tools for Modeling Catalysis. *Israel Journal of Chemistry* **2022,** *62* (1-2), e202100136.
29. Mardirossian, N.; Head-Gordon, M., Mapping the genome of meta-generalized gradient approximation density functionals: The search for B97M-V. *The Journal of Chemical Physics* **2015,** *142* (7), 074111.
30. Epifanovsky, E.; Gilbert, A. T. B.; Feng, X.; Lee, J.; Mao, Y.; Mardirossian, N.; Pokhilko, P.; White, A. F.; Coons, M. P.; Dempwolff, A. L.; Gan, Z.; Hait, D.; Horn, P. R.; Jacobson, L. D.; Kaliman, I.; Kussmann, J.; Lange, A. W.; Lao, K. U.; Levine, D. S.; Liu, J.; McKenzie, S. C.; Morrison, A. F.; Nanda, K. D.; Plasser, F.; Rehn, D. R.; Vidal, M. L.; You, Z.-Q.; Zhu, Y.; Alam, B.; Albrecht, B. J.; Aldossary, A.; Alguire, E.; Andersen, J. H.; Athavale, V.; Barton, D.; Begam, K.; Behn, A.; Bellonzi, N.; Bernard, Y. A.; Berquist, E. J.; Burton, H. G. A.; Carreras, A.; Carter-Fenk, K.; Chakraborty, R.; Chien, A. D.; Closser, K. D.; Cofer-Shabica, V.; Dasgupta, S.; de Wergifosse, M.; Deng, J.; Diedenhofen, M.; Do, H.; Ehlert, S.; Fang, P.-T.; Fatehi, S.; Feng, Q.; Friedhoff, T.; Gayvert, J.; Ge, Q.; Gidofalvi, G.; Goldey, M.; Gomes, J.; González-Espinoza, C. E.; Gulania, S.; Gunina, A. O.; Hanson-Heine, M. W. D.; Harbach, P. H. P.; Hauser, A.; Herbst, M. F.; Hernández Vera, M.; Hodecker, M.; Holden, Z. C.; Houck, S.; Huang, X.; Hui, K.; Huynh, B. C.; Ivanov, M.; Jász, Á.; Ji, H.; Jiang, H.; Kaduk, B.; Kähler, S.; Khistyaev, K.; Kim, J.; Kis, G.; Klunzinger, P.; Koczor-Benda, Z.; Koh, J. H.; Kosenkov, D.; Koulias, L.; Kowalczyk, T.; Krauter, C. M.; Kue, K.; Kunitsa, A.; Kus, T.; Ladjánszki, I.; Landau, A.; Lawler, K. V.; Lefrancois, D.; Lehtola, S.; Li, R. R.; Li, Y.-P.; Liang, J.; Liebenthal, M.; Lin, H.-H.; Lin, Y.-S.; Liu, F.; Liu, K.-Y.; Loipersberger, M.; Luenser, A.; Manjanath, A.; Manohar, P.; Mansoor, E.; Manzer, S. F.; Mao, S.-P.; Marenich, A. V.; Markovich, T.; Mason, S.; Maurer, S. A.; McLaughlin, P. F.; Menger, M. F. S. J.; Mewes, J.-M.; Mewes, S. A.; Morgante, P.; Mullinax, J. W.; Oosterbaan, K. J.; Paran, G.; Paul, A. C.; Paul, S. K.; Pavošević, F.; Pei, Z.; Prager, S.; Proynov, E. I.; Rák, Á.; Ramos-Cordoba, E.; Rana, B.; Rask, A. E.; Rettig, A.; Richard, R. M.; Rob, F.; Rossomme, E.; Scheele, T.; Scheurer, M.; Schneider, M.; Sergueev, N.; Sharada, S. M.; Skomorowski, W.; Small, D. W.; Stein, C. J.; Su, Y.-C.; Sundstrom, E. J.; Tao, Z.; Thirman, J.; Tornai, G. J.; Tsuchimochi, T.; Tubman, N. M.; Veccham, S. P.; Vydrov, O.; Wenzel, J.; Witte, J.; Yamada, A.; Yao, K.; Yeganeh, S.; Yost, S. R.; Zech, A.; Zhang, I. Y.; Zhang, X.; Zhang, Y.; Zuev, D.; Aspuru-Guzik, A.; Bell, A. T.; Besley, N. A.; Bravaya, K. B.; Brooks, B. R.; Casanova, D.; Chai, J.-D.; Coriani, S.; Cramer, C. J.; Cserey, G.; DePrince, A. E.; DiStasio, R. A.; Dreuw, A.; Dunietz, B. D.; Furlani, T. R.; Goddard, W. A.; Hammes-Schiffer, S.; Head-Gordon, T.; Hehre, W. J.; Hsu, C.-P.; Jagau, T.-C.; Jung, Y.; Klamt, A.; Kong, J.; Lambrecht, D. S.; Liang, W.; Mayhall, N. J.; McCurdy, C. W.; Neaton, J. B.; Ochsenfeld, C.; Parkhill, J. A.; Peverati, R.; Rassolov, V. A.; Shao, Y.; Slipchenko, L. V.; Stauch, T.; Steele, R. P.; Subotnik, J. E.; Thom, A. J. W.; Tkatchenko, A.; Truhlar, D. G.; Van Voorhis, T.; Wesolowski, T. A.; Whaley, K. B.; Woodcock, H. L.; Zimmerman, P. M.; Faraji, S.; Gill, P. M. W.; Head-Gordon, M.; Herbert, J. M.; Krylov, A. I., Software for the frontiers of quantum chemistry: An overview of developments in the Q-Chem 5 package. *The Journal of Chemical Physics* **2021,** *155* (8), 084801.

31. Kühne, T. D.; Iannuzzi, M.; Del Ben, M.; Rybkin, V. V.; Seewald, P.; Stein, F.; Laino, T.; Khaliullin, R. Z.; Schütt, O.; Schiffmann, F.; Golze, D.; Wilhelm, J.; Chulkov, S.; Bani-Hashemian, M. H.; Weber, V.; Borštnik, U.; Taillefumier, M.; Jakobovits, A. S.; Lazzaro, A.; Pabst, H.; Müller, T.; Schade, R.; Guidon, M.; Andermatt, S.; Holmberg, N.; Schenter, G. K.; Hehn, A.; Bussy, A.; Belleflamme, F.; Tabacchi, G.; Glöß, A.; Lass, M.; Bethune, I.; Mundy, C. J.; Plessl, C.; Watkins, M.; VandeVondele, J.; Krack, M.; Hutter, J. C. K., An Electronic Structure and Molecular Dynamics Software Package - Quickstep: Efficient and Accurate Electronic Structure Calculations. *J. Chem. Phys* **2020,** *152* (19), 194103-194103.
32. Mardirossian, N.; Ruiz Pestana, L.; Womack, J. C.; Skylaris, C.-K.; Head-Gordon, T.; Head-Gordon, M., Use of the rVV10 Nonlocal Correlation Functional in the B97M-V Density Functional: Defining B97M-rV and Related Functionals. *The Journal of Physical Chemistry Letters* **2017,** *8* (1), 35-40.
33. Goerigk, L.; Grimme, S., A thorough benchmark of density functional methods for general main group thermochemistry, kinetics, and noncovalent interactions. *Physical Chemistry Chemical Physics* **2011,** *13* (14), 6670-6688.
34. Becke, A. D., Density-functional exchange-energy approximation with correct asymptotic behavior. *Physical Review A* **1988,** *38* (6), 3098-3100.
35. Perdew, J. P.; Burke, K.; Ernzerhof, M., Generalized gradient approximation made simple. *Phys. Rev. Lett.* **1996,** *77* (18), 3865.
36. Li, W. L.; Chen, K.; Rossomme, E.; Head-Gordon, M.; Head-Gordon, T., Greater transferability and accuracy of norm-conserving pseudopotentials using nonlinear core corrections. *Chem Sci* **2023,** *14* (39), 10934-10943.
37. Kendelewicz, T.; Kaya, S.; Newberg, J. T.; Bluhm, H.; Mulakaluri, N.; Moritz, W.; Scheffler, M.; Nilsson, A.; Pentcheva, R.; Brown, G. E., Jr., X-ray Photoemission and Density Functional Theory Study of the Interaction of Water Vapor with the Fe3O4(001) Surface at Near-Ambient Conditions. *The Journal of Physical Chemistry C* **2013,** *117* (6), 2719-2733.
38. Naveas, N.; Pulido, R.; Marini, C.; Gargiani, P.; Hernandez-Montelongo, J.; Brito, I.; Manso-Silván, M., First-Principles Calculations of Magnetite (Fe3O4) above the Verwey Temperature by Using Self-Consistent DFT + U + V. *Journal of Chemical Theory and Computation* **2023,** *19* (23), 8610-8623.
39. Chan, B.; Gill, P. M. W.; Kimura, M., Assessment of DFT Methods for Transition Metals with the TMC151 Compilation of Data Sets and Comparison with Accuracies for Main-Group Chemistry. *Journal of Chemical Theory and Computation* **2019,** *15* (6), 3610-3622.
40. Wappett, D. A.; Goerigk, L., Benchmarking Density Functional Theory Methods for Metalloenzyme Reactions: The Introduction of the MME55 Set. *Journal of Chemical Theory and Computation* **2023,** *19* (22), 8365-8383.
41. Cornell, R. M.; Schwertman, U., The Iron Oxides: Structure, Properties, Reactions, Occurrence and Uses **1997,** *15* (3-4), 533-559.
42. Goodacre, D.; Blum, M.; Buechner, C.; Hoek, H.; Gericke, S. M.; Jovic, V.; Franklin, J. B.; Kittiwatanakul, S.; Söhnel, T.; Bluhm, H.; Smith, K. E., Water adsorption on vanadium oxide thin films in ambient relative humidity. *The Journal of Chemical Physics* **2020,** *152* (4), 044715.
43. Degaga, G. D.; Trought, M.; Nemsak, S.; Crumlin, E. J.; Seel, M.; Pandey, R.; Perrine, K. A., Investigation of N(2) adsorption on Fe(3)O(4)(001) using ambient pressure X-ray photoelectron spectroscopy and density functional theory. *J Chem Phys* **2020,** *152* (5), 054717.
44. Murakami, J.; Yamaguchi, W., Reduction of N2 by supported tungsten clusters gives a model of the process by nitrogenase. *Sci Rep* **2012,** *2*, 407.
45. Alberas, D. J.; Kiss, J.; Liu, Z. M.; White, J. M., Surface-Chemistry of Hydrazine on Pt(111). *Surface Science* **1992,** *278* (1-2), 51-61.

46. Apen, E.; Gland, J. L., Hydrazine adsorption and decomposition on the GaAs(100)-c(8 × 2) surface. *Surface Science* **1994,** *321* (3), 308-317.
47. Bu, Y.; Lin, M. C., Surface chemistry of N2H4 on Si(100)-2 × 1. *Surface Science* **1994,** *311* (3), 385-394.
48. Grunze, M., The interaction of hydrazine with an Fe(111) surface. *Surface Science* **1979,** *81* (2), 603-625.

# Supplementary Information
# Ammonia Synthesis under Ambient Conditions: Insights into Water-Nitrogen-Magnetite Interfaces

Sruthy K. Chandy[1,2], Mauricio Lopez Luna[2], Nykita Z. Rustad[3], Isaac N. Zakaria[3,5], Andreas Siebert[2,3], Shane Devlin[3,6], Wan-Lu Li[7, 8], Monika Blum[2,3]*, Teresa Head-Gordon[1,2,4,5*]

[1]Kenneth S. Pitzer Theory Center and Department of Chemistry, University of California, Berkeley, CA, 94720 USA
[2]Chemical Sciences Division, Lawrence Berkeley National Laboratory, Berkeley, CA, 94720 USA
[3]Advanced Light Source, Lawrence Berkeley National Laboratory, Berkeley, CA, 94720 USA
[4]Departments of Bioengineering and [5]Chemical and Biomolecular Engineering, University of California, Berkeley, CA, 94720, USA
[6]Nevada Extreme Conditions Laboratory, University of Nevada, Las Vegas, NV, 89154, USA
[7]Aiiso Yufeng Li Family Department of Chemical and Nano Engineering and [8]Materials Science and Engineering, University of California, San Diego, La Jolla, CA 92093 USA

## Supplementary Methods

Fitting Procedure and Uncertainties: All XPS data were analyzed using CasaXPS,[1] with peak fitting performed on the O 1s, Fe 2p, and Fe 3p spectra. A Shirley background correction was applied to the O 1s, Fe 2p, and Fe 3p spectra, and the N 1s spectrum was fitted with a linear background. For the O 1s spectra, the Fe-O peak was modeled using a GL(60)T(2) line shape, i.e., a convoluted Gaussian peak with an asymmetry factor, while the remaining peaks were fitted using a LA(1.643) line shape. The binding energy assignments were as follows: Fe-O at 529.9 eV, C-O at 531.2 eV, -OH at 531.7 eV, liquid-phase $H_2O$ at 532.8 eV, and gaseous-phase $H_2O$ at 534.8 eV.[2-3] The Fe 2p spectra were deconvoluted into two sets of doublets along with corresponding satellite features. The $Fe^{3+}$ and $Fe^{2+}$ states were fitted using GL(60)T(0.65) for the $2p_{3/2}$ components and GL(90)T(1.15) for the $2p_{1/2}$ components. The $Fe^{3+}$ $2p_{3/2}$ peak was positioned at 711.0 eV, with the $2p_{1/2}$ counterpart set 13.7 eV higher. Similarly, the $Fe^{2+}$ $2p_{3/2}$ peak was assigned to 709.5 eV, maintaining a 13.7 eV separation for the $2p_{1/2}$ peak.[4-6] The Fe 3p spectrum was recorded using a photon energy of 255 eV and fitted with $Fe^{3+}$ and $Fe^{2+}$ components. The $Fe^{3+}$ peak was referenced to 56 eV, and this reference was used to calibrate all spectra to ensure consistency and accuracy in binding energy assignments. The N 1s spectrum was fitted using an LA(1.643) line shape, with three peaks assigned to $N_2H_2$ at 401.1 eV, $NH_3$ at 399.9 eV, and $Fe_0^{3+}$-N at 399.0 eV.[7-11]

Uncertainty in the fitted peak areas was quantified by extracting the standard deviation (σ) from the covariance matrix in CasaXPS, where σ corresponds to the square root of the relevant area diagonal element. The propagation of uncertainty for peak area ratios ($R=A_1/A_2$) was determined using:

$$\sigma_R = R\sqrt{\left(\frac{\sigma_{A_1}}{A_1}\right)^2 + \left(\frac{\sigma_{A_2}}{A_2}\right)^2}$$

where $A_1$ and $A_2$ are the respective peak areas, and $\sigma_{A_1}$ and $\sigma_{A_2}$ are their uncertainties. The uncertainty in temperature measurements was estimated as the standard deviation of the recorded temperatures obtained across multiple measurements under identical conditions.

## Supplementary Tables

**Table S1:** The relative energies of M15, M5 and M7 species with respect to M15 species for a) and b) basis sets investigated using 5 different DFT functionals. All energies in eV. See Figure S3.

| **Basis Set** | DZVP-MOLOPT | | | | |
|---|---|---|---|---|---|
| **Species/Functional** | **PBE** | **BP86** | **B97M-rV** | **revPBE0-D3** | **wB97X-V** |
| M15 | 0 | 0 | 0 | 0 | 0 |
| M5 | -0.14 | 0.22 | 0.11 | 0.05 | 0.01 |
| M7 | -0.02 | 0.68 | 0.34 | 0.02 | -0.04 |
| **Basis Set** | TZVP-MOLOPT | | | | |
| **Species/Functional** | **PBE** | **BP86** | **B97M-rV** | **revPBE0-D3** | **wB97X-V** |
| M15 | 0 | 0 | 0 | 0 | 0 |
| M5 | -0.12 | 0.02 | 0.16 | 0.05 | 0.01 |
| M7 | 0.004 | -0.02 | 0.37 | 0.03 | -0.04 |

# Supplementary Figures

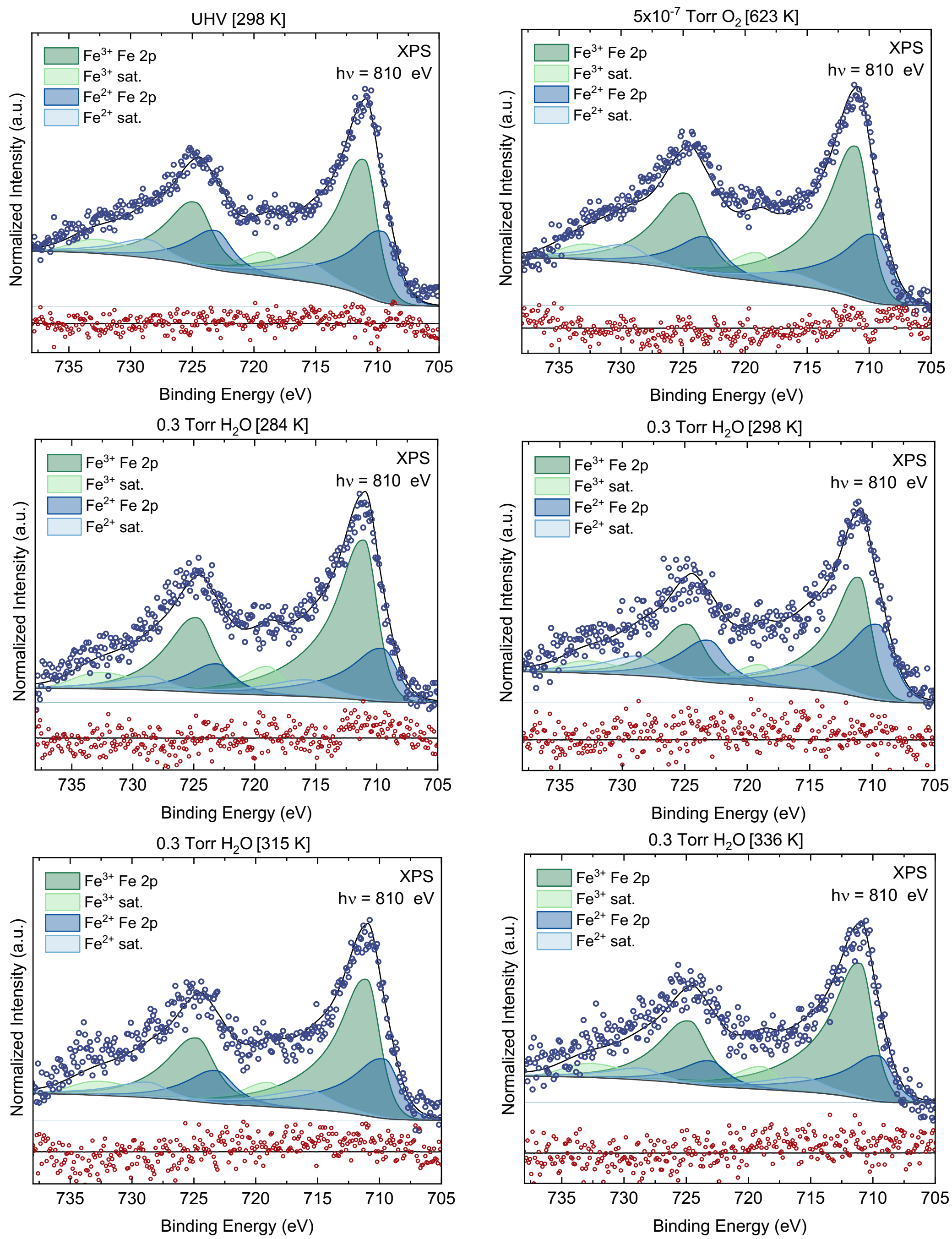

**Figure S1**. *APXPS peak fitting of the Fe 2p spectra of the $Fe_3O_4$ nanoparticles exposed to 0.3 Torr of $H_2O$ as a function of temperature.*

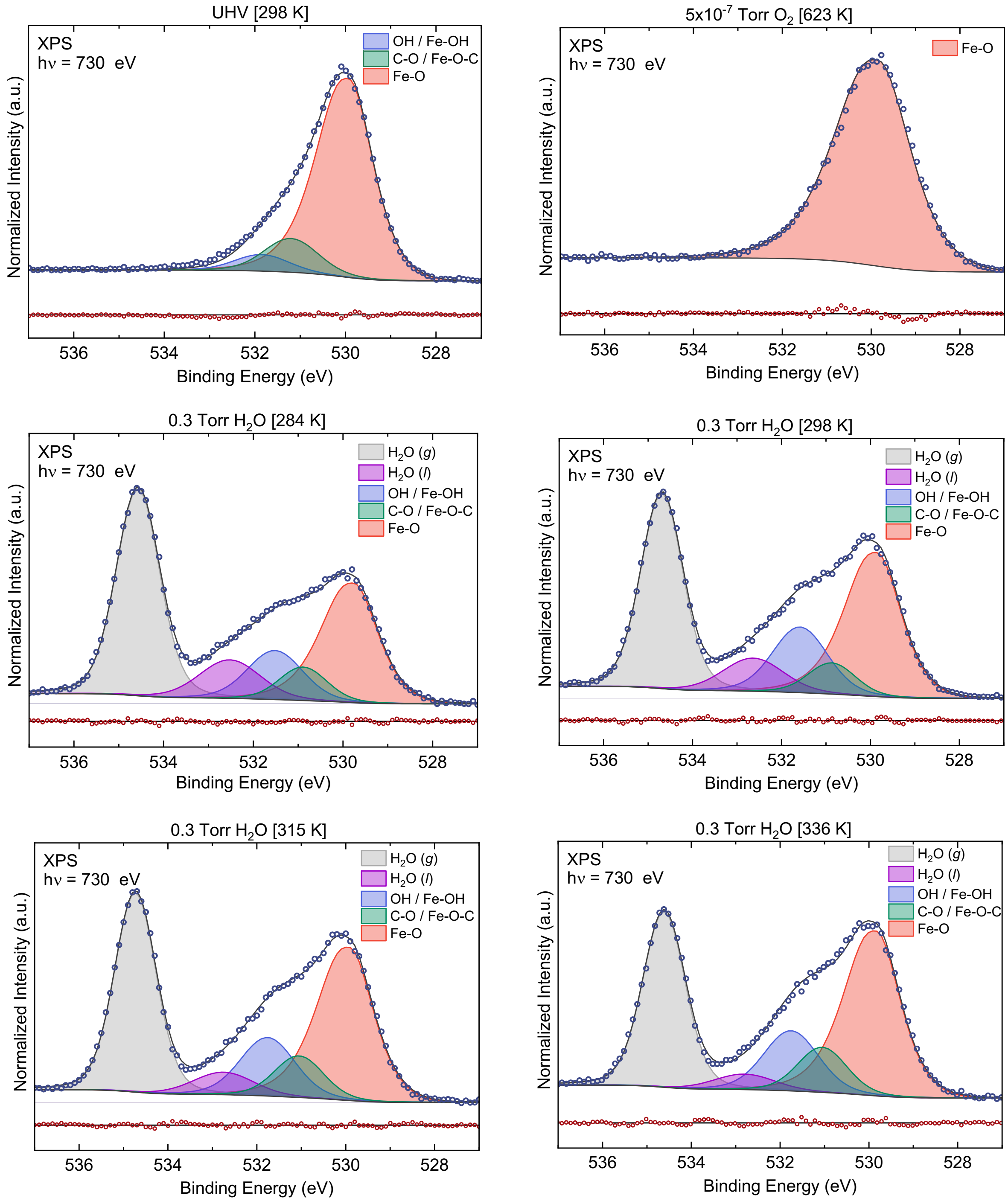

**Figure S2**. *APXPS peak fitting of the O 1s region of the* $Fe_3O_4$ *nanoparticles exposed to 0.3 Torr of* $H_2O$ *as a function of temperature.*

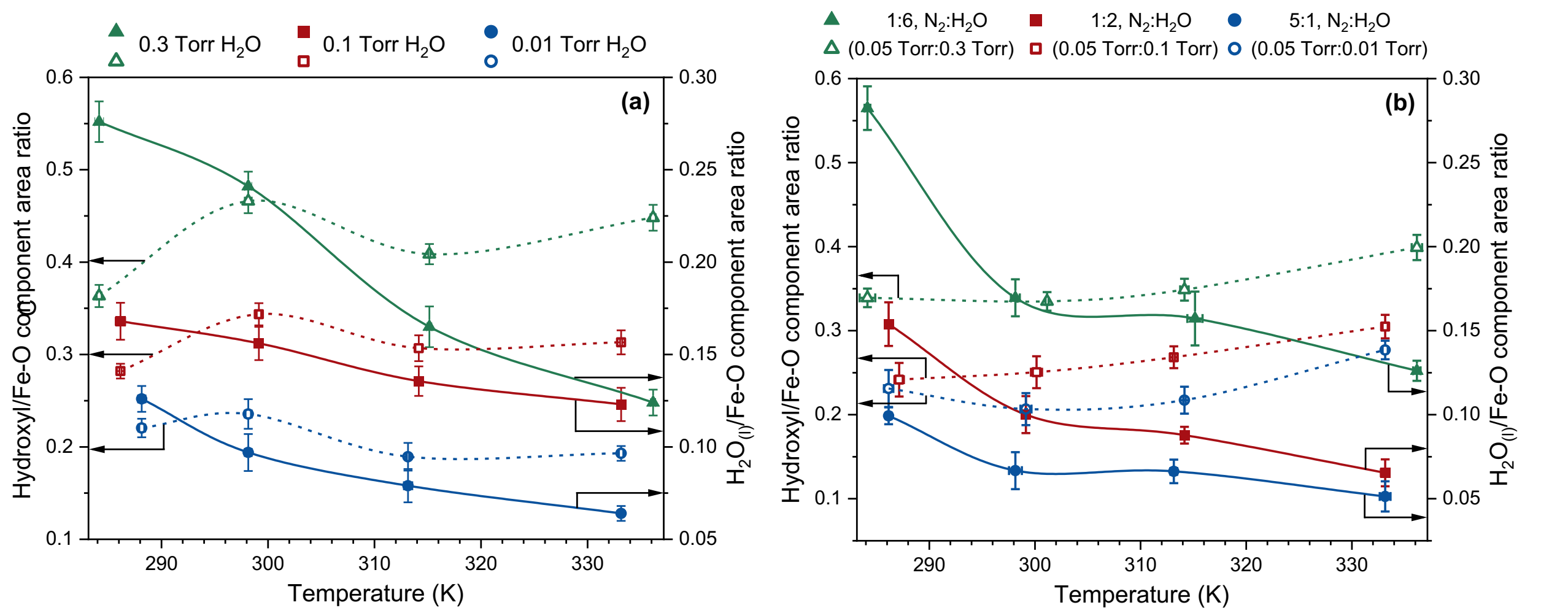

**Figure S3**. *APXPS core level spectra of the $Fe_3O_4$ nanoparticles when exposed to water and nitrogen as a function of temperature.* Experimental component area ratios for $OH^-$/Fe-O (open symbols) and $H_2O$(l)/Fe-O (closed symbols) as a function of temperature at different pressures (a) $H_2O$ vapor only and (b) with $N_2$ gas and $H_2O$ vapor

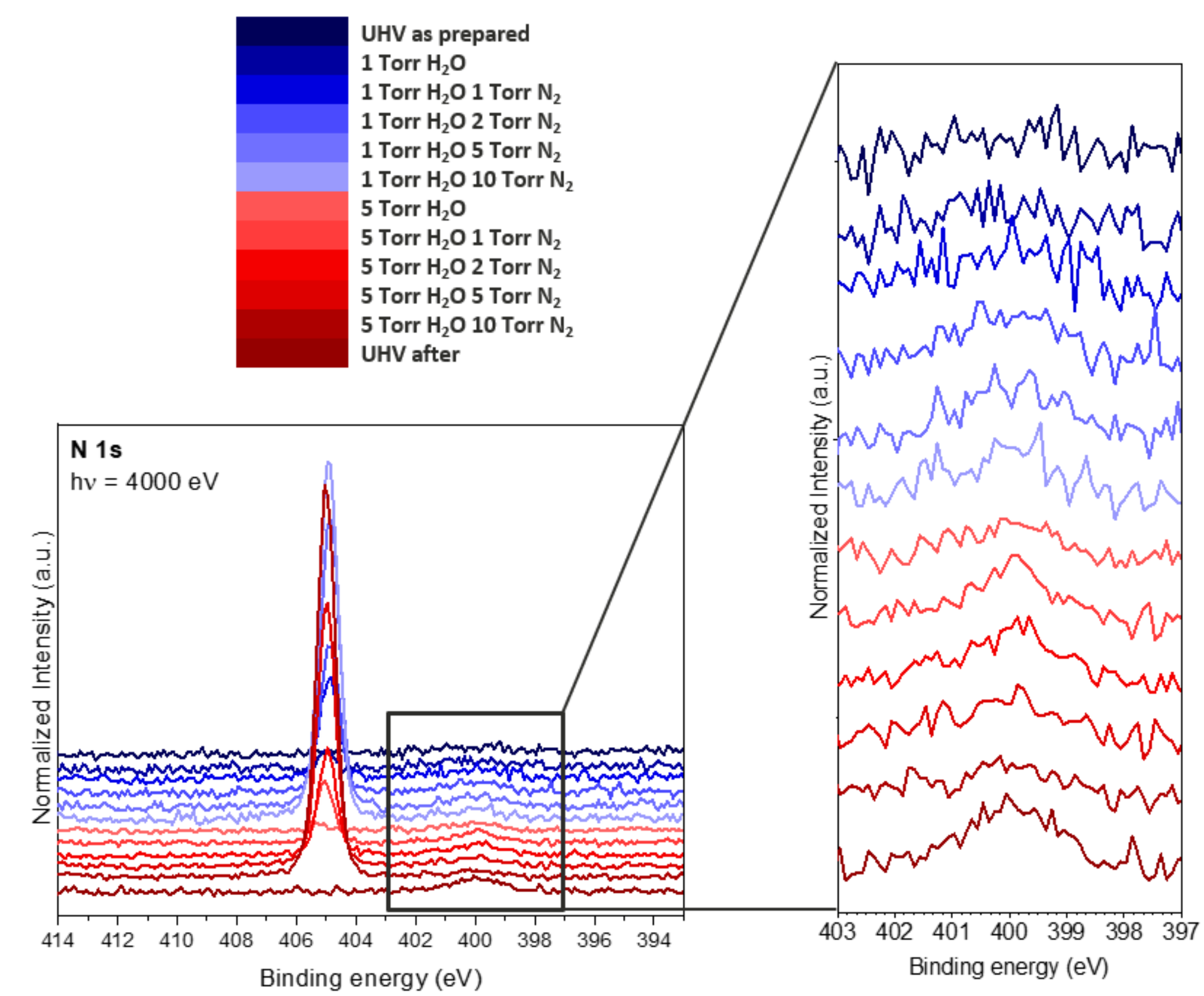

**Figure S4**. *APXPS core level spectra using tender X-rays to monitor the N 1s of the $Fe_3O_4$ nanoparticles when exposed to $N_2$ gas and $H_2O$ vapor as a function of gas pressures*. Experiments reveal new nitrogen species, which are analyzed in the main text.

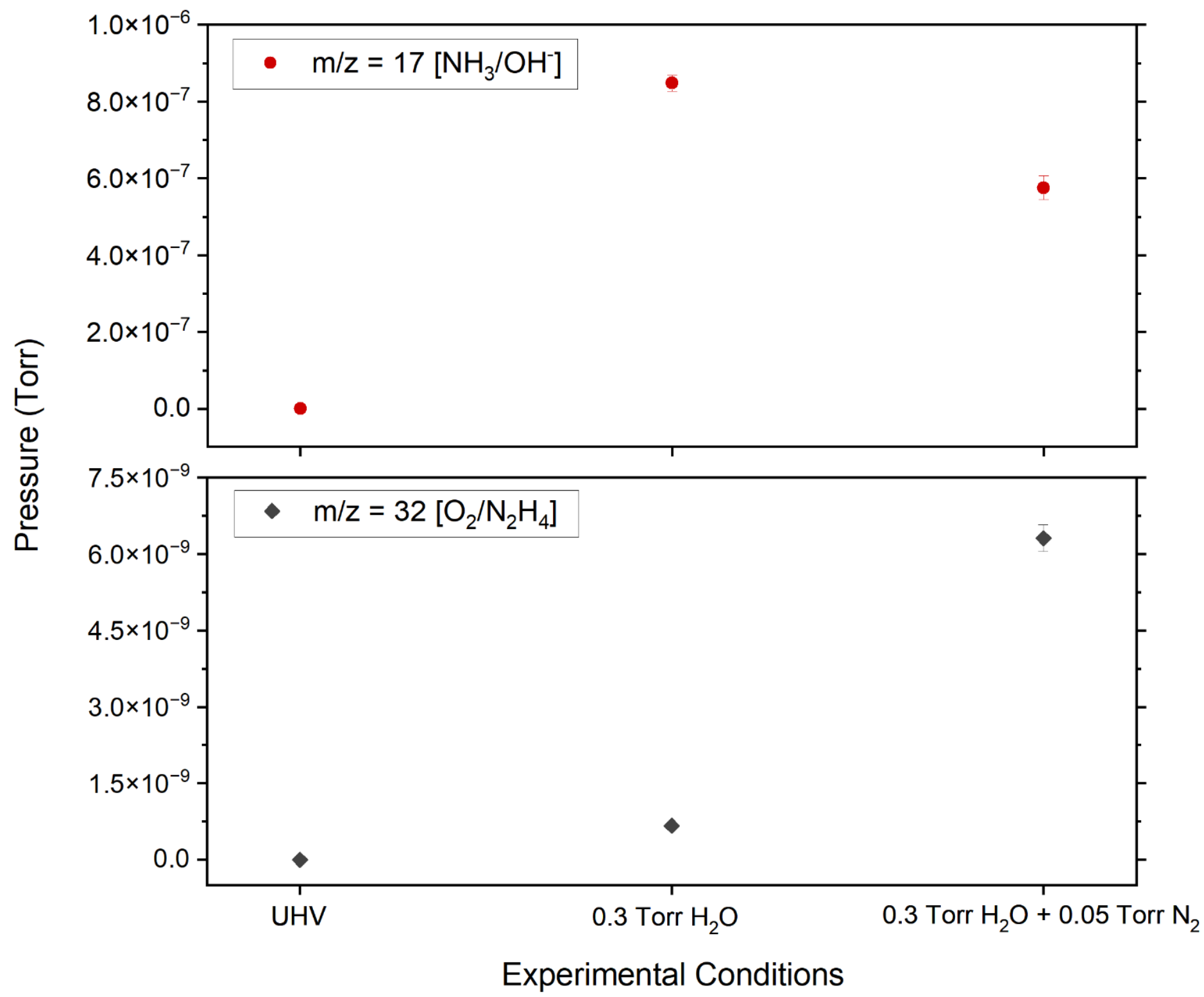

**Figure S5.** *Mass spectroscopy data. for m/z 17 (top) and m/z 32 (bottom).* Experimental conditions at UHV (RT), exposed to 0.3 Torr of $H_2O$ (RT), and 0.3 Torr of $H_2O$ with 0.05 Torr of $N_2$ gas (RT).

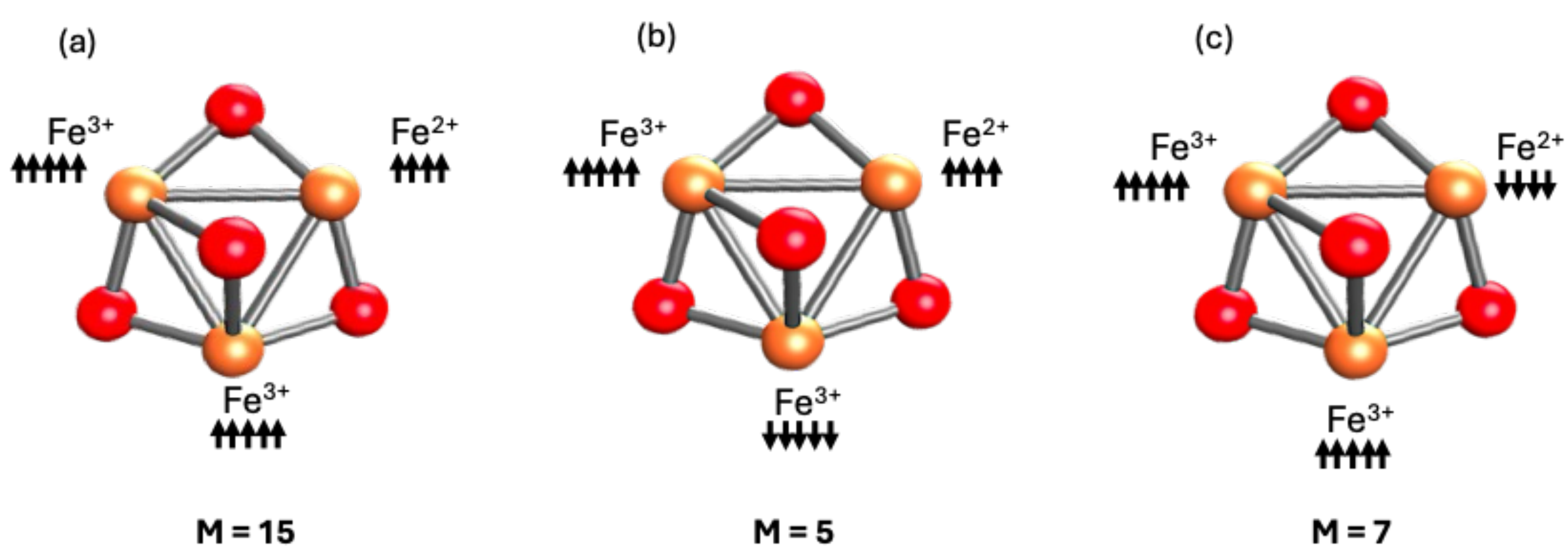

**Figure S6:** *Different spin states of $Fe_3O_4$ nanoparticles considered in this study.*

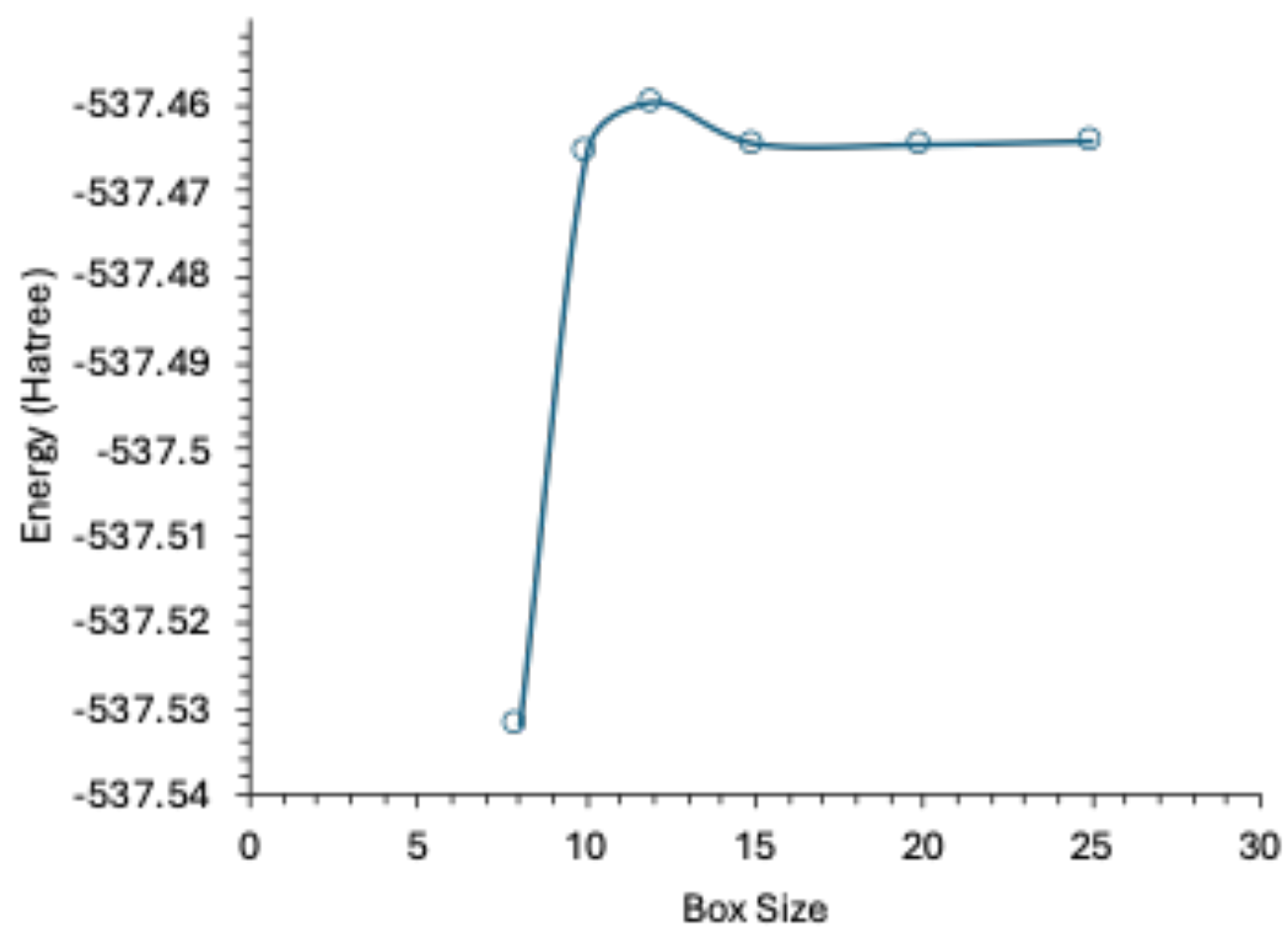

**Figure S7**: Benchmarking the box size for AIMD simulations with box size on x axis in angstrom and the single point energy in hartrees for the Fe3O4 with six water clusters on y axis.

(a) (b)

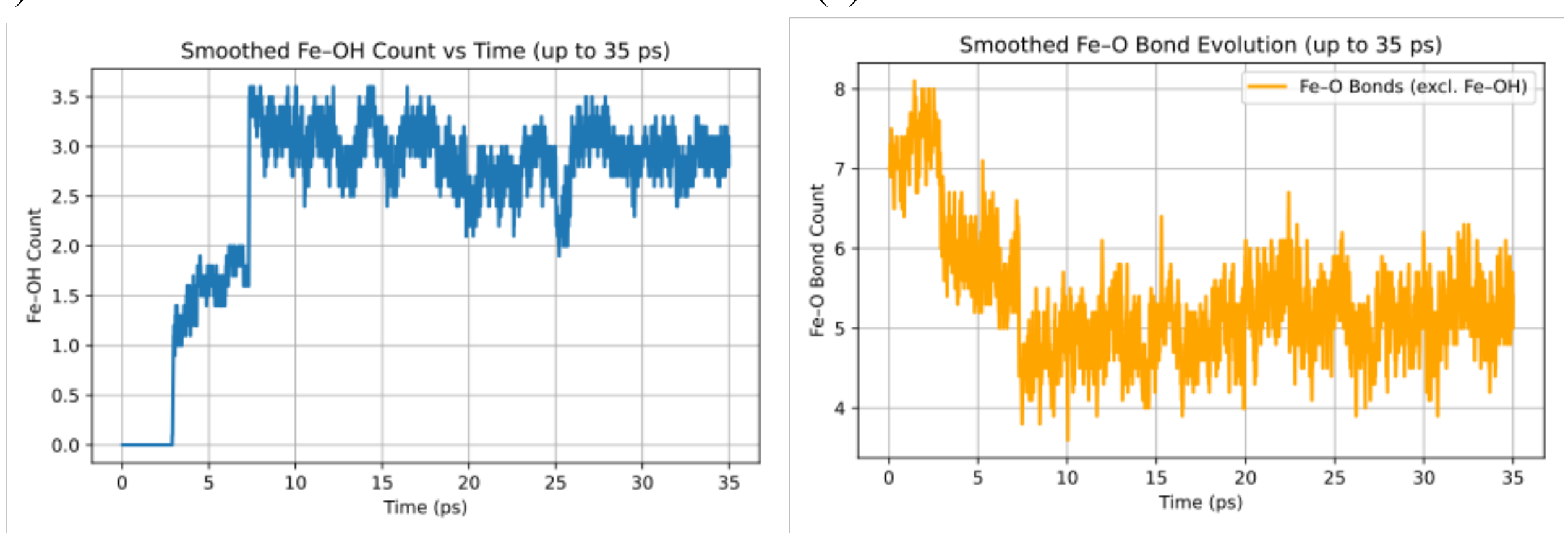

**Figure S8.** Time evolution of (a) Fe–OH groups and (b) intact Fe–O bonds during a 35 ps *ab initio* molecular dynamics simulation of $Fe_3O_4$ with six interfacial water molecules at 298 K. Fe–OH and intact Fe–O groups were identified using distance cutoffs derived from optimized B97M-rV/TZVP-MOLOPT geometries of hydroxylated $Fe_3O_4$ species: 0.96 Å for the O–H bond and 1.96 Å for the Fe–O bond. To account for thermal fluctuations, we applied a ±0.05 Å buffer, resulting in thresholds of ≤1.1 Å for O–H and ≤2.1 Å for Fe–O.

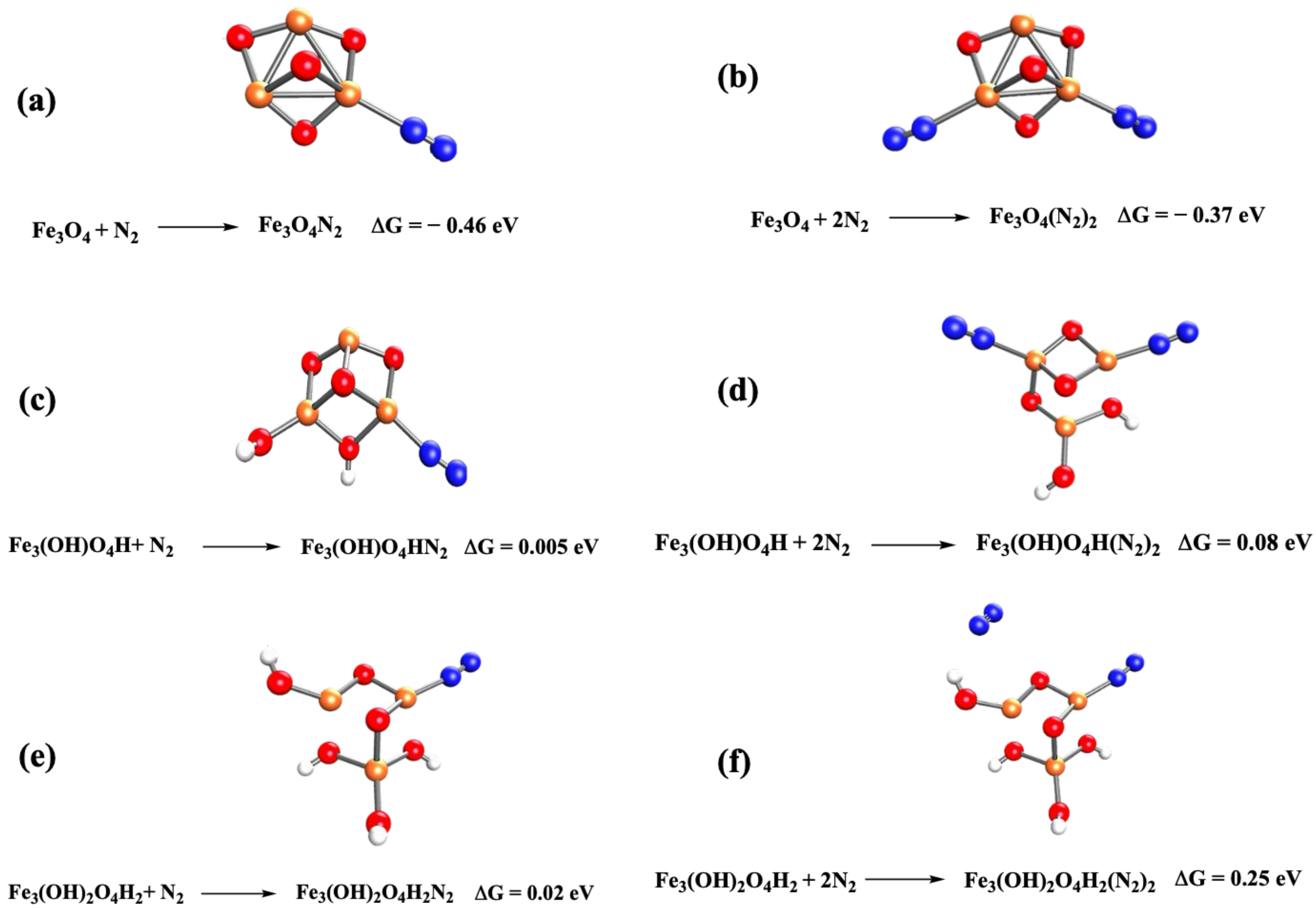

**Figure S9:** *Free energy change for increasing nitrogen addition to the iron oxide nanoparticle.* (a,b) the bare $Fe_3O_4$, and increasing water splitting for (c,d) $Fe_3(OH)O_4(H)$ and (e,f) $Fe_3(OH)_2O_4(H)_2$

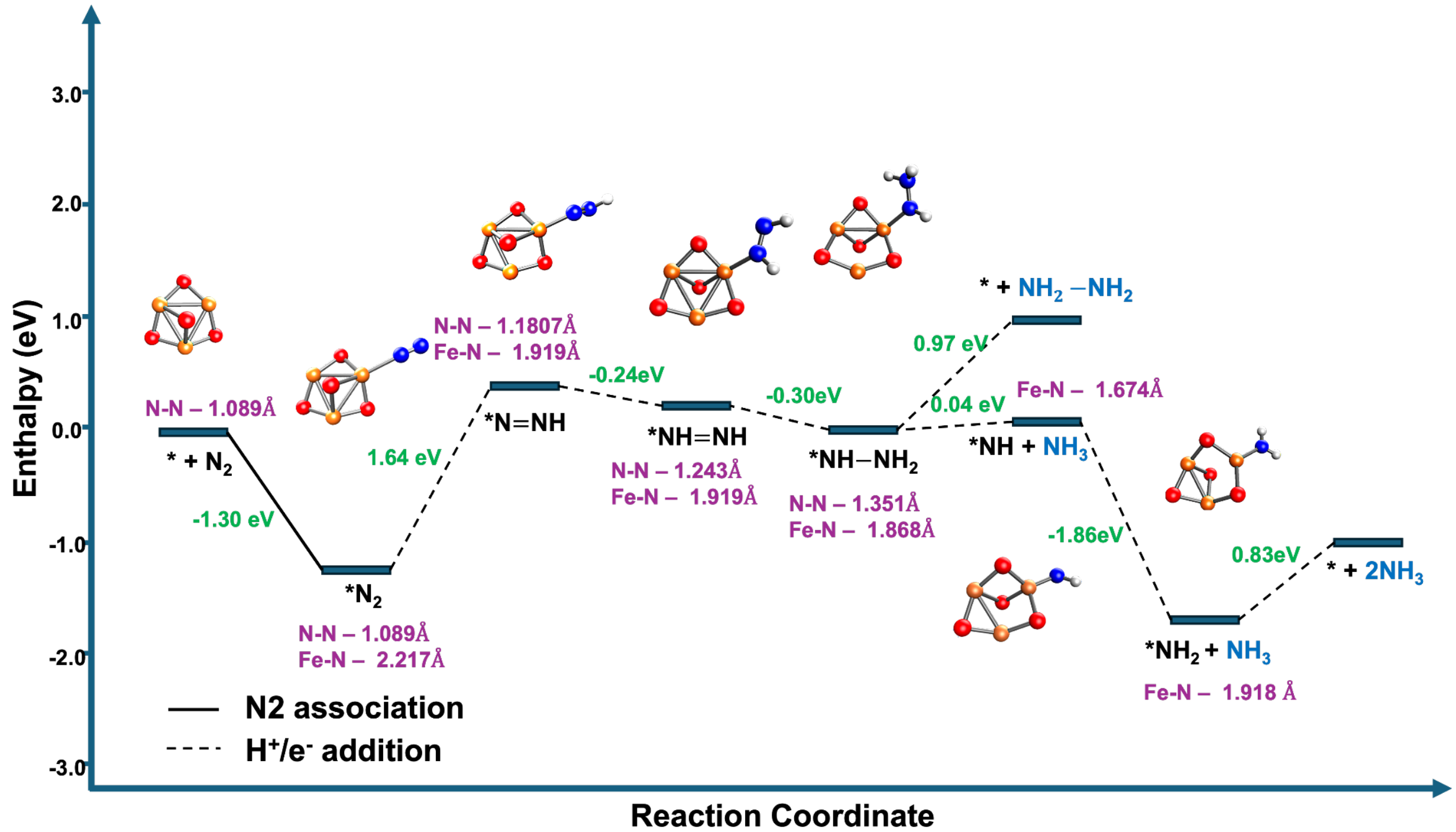

**Figure S10:** *Enthalpy change for the NRR mechanism on bare $Fe_3O_4$.* * refers to bare $Fe_3O_4$ surface (no water dissociation).

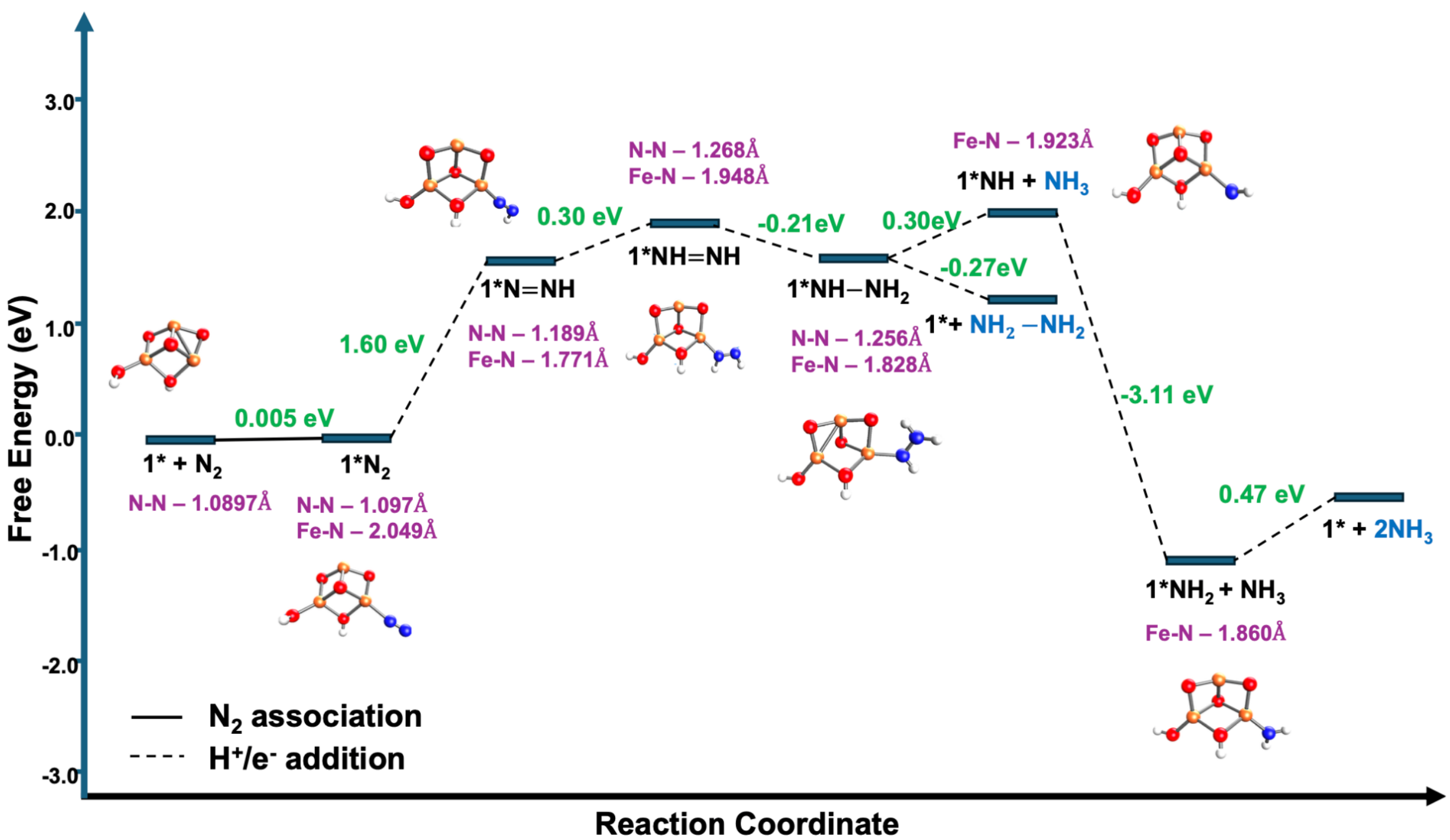

**Figure S11:** *Free energy change for the NRR mechanism on $Fe_3O_4$ after one water dissociation*. 1* refers to $Fe_3(OH)O_4(H)$.

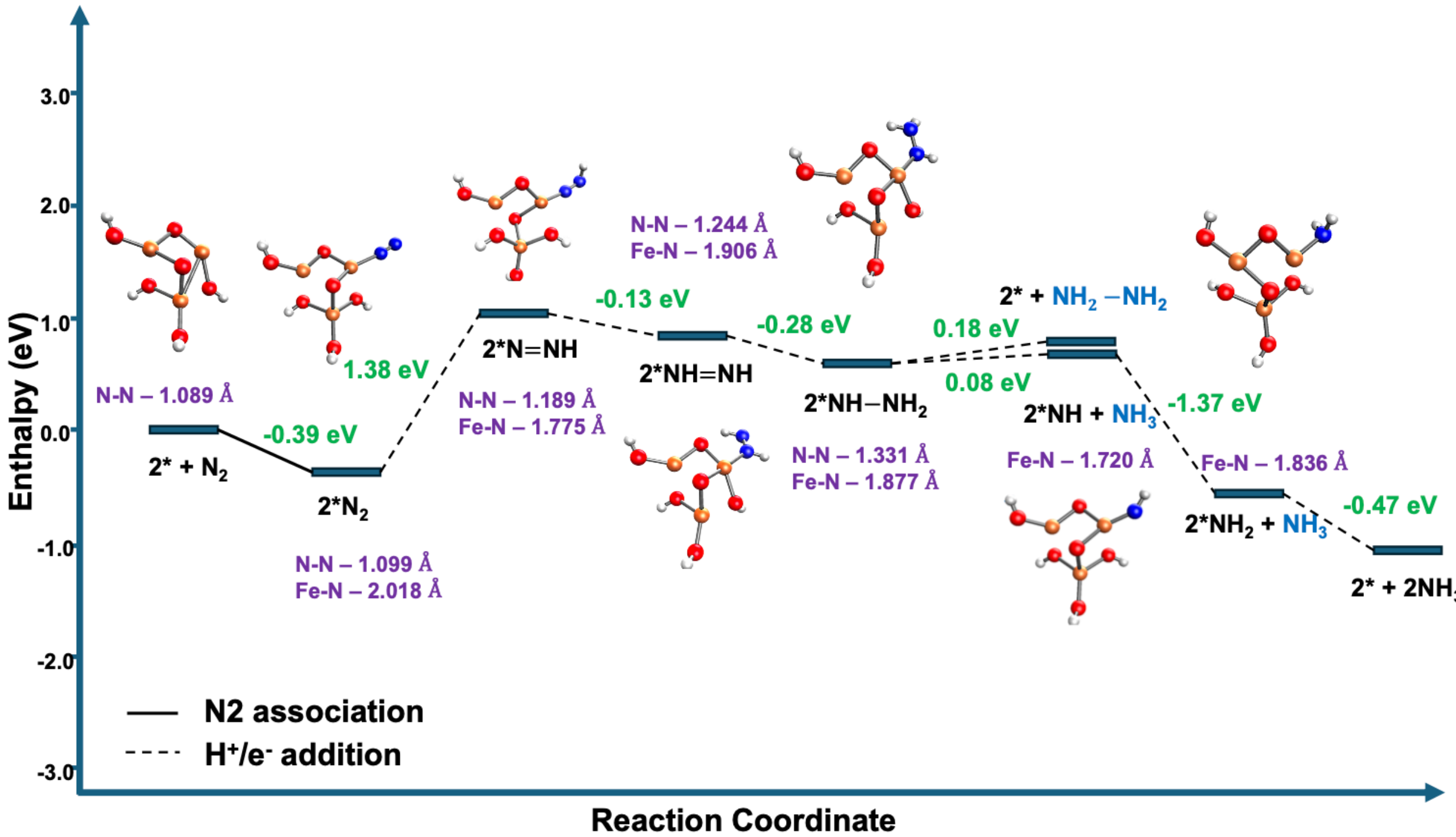

**Figure S12:** *Enthalpy change for the NRR mechanism on $Fe_3O_4$ after two water dissociation.* 2* refers to $Fe_3(OH)_2O_4(H)_2$

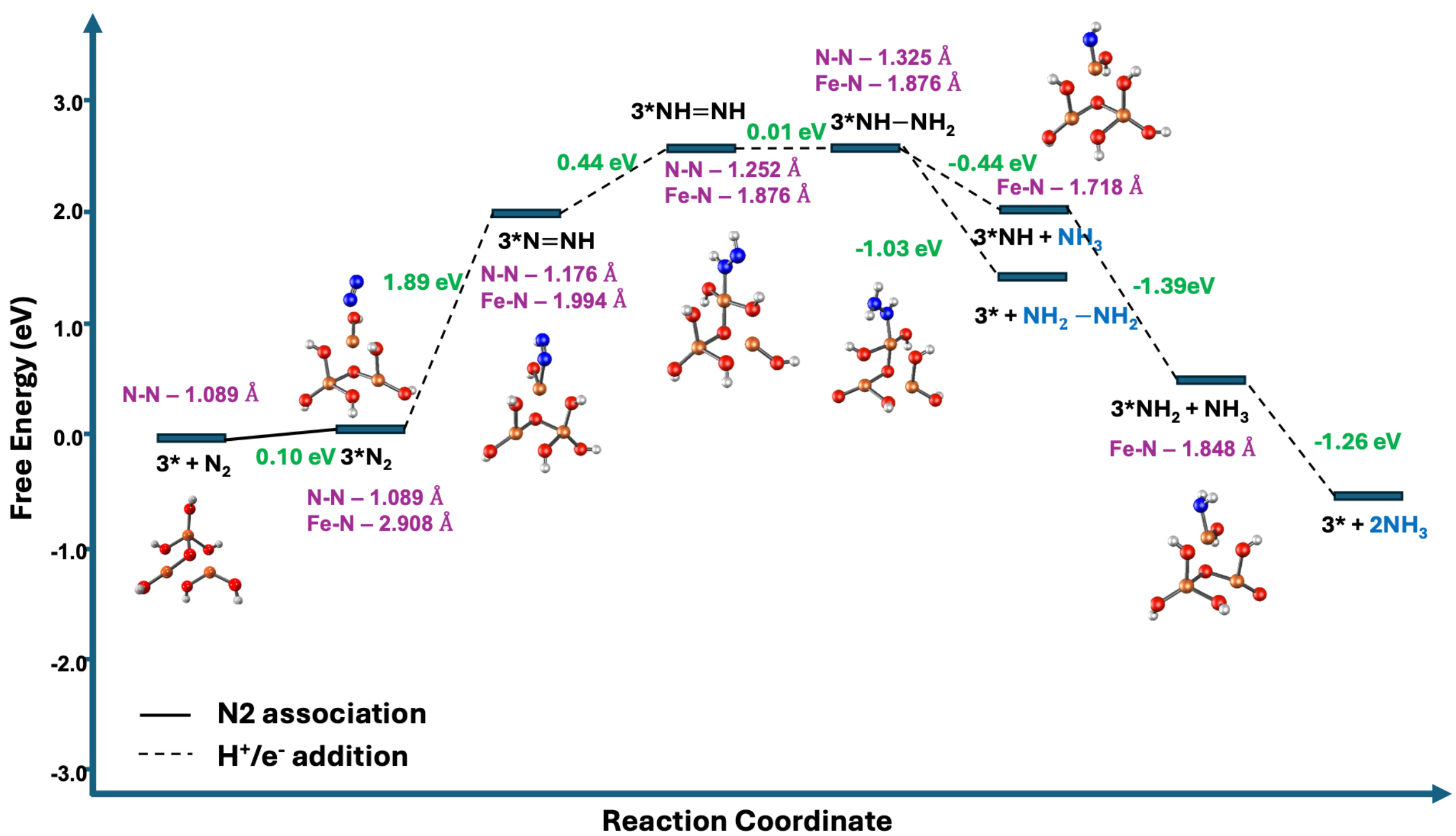

**Figure S13:** *Free energy change for the NRR mechanism on $Fe_3O_4$ after three water dissociations.* 3* refers to $Fe_3(OH)_3O_4(H)_3$.

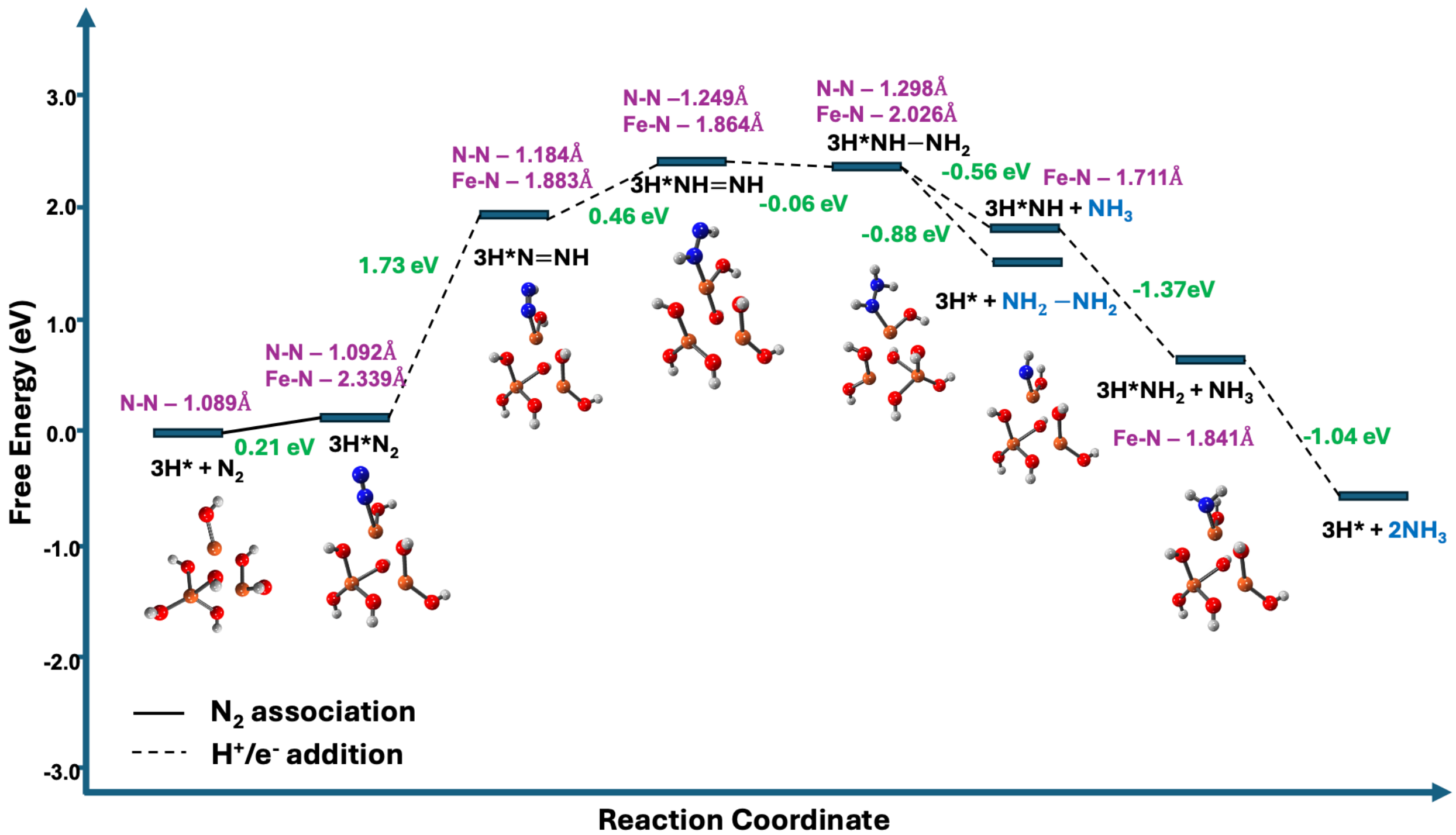

**Figure S14:** *Free energy change for the NRR mechanism on $Fe_3O_4$ after three water dissociations plus additional hydrogen/electron*. 3H* refers to $Fe_3(OH)_3O_4(H)_4$.

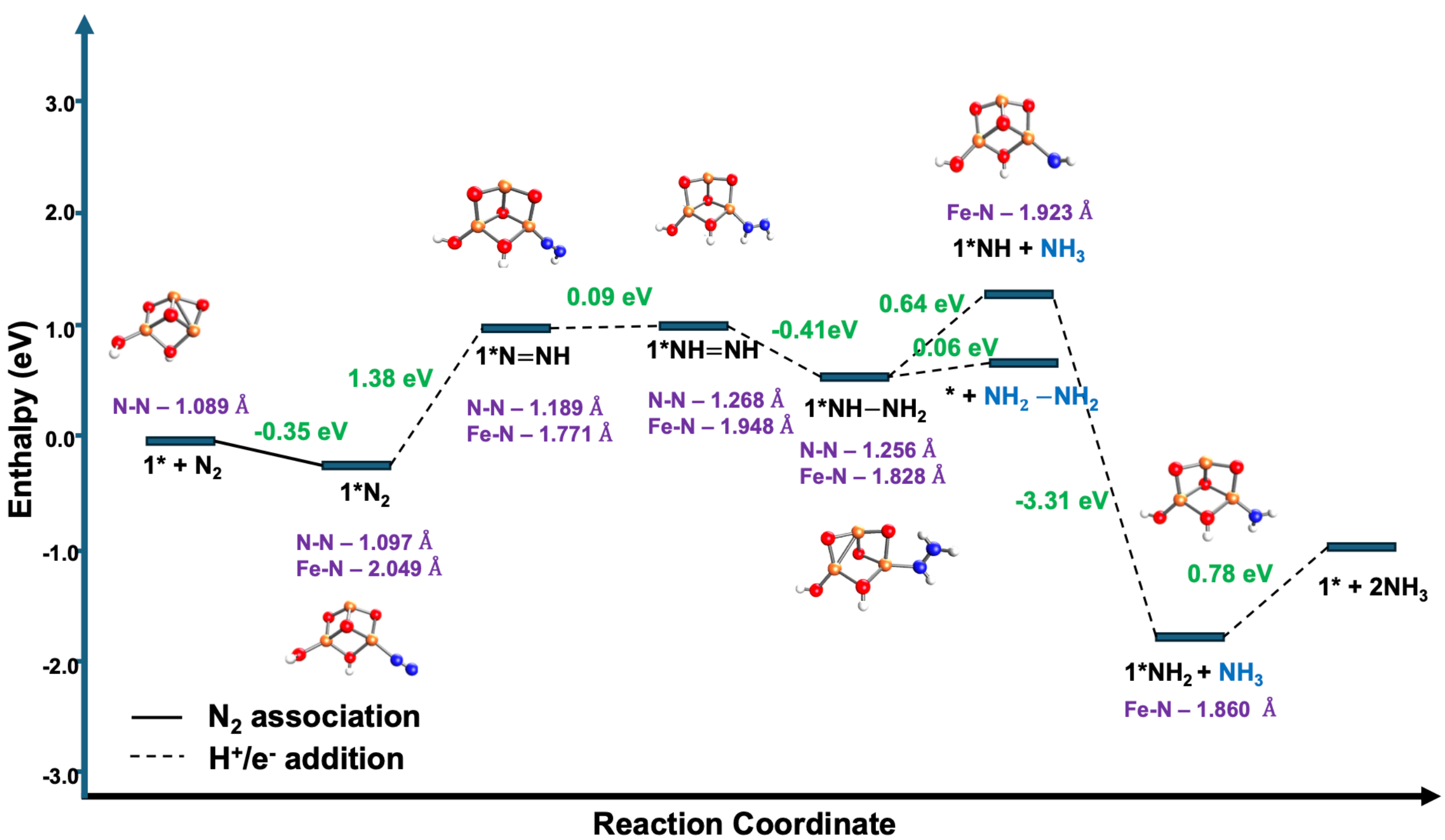

**Figure S15:** *Enthalpy change for the NRR mechanism on $Fe_3O_4$ after one water dissociation.* 1* refers to $Fe_3(OH)O_4(H)$.

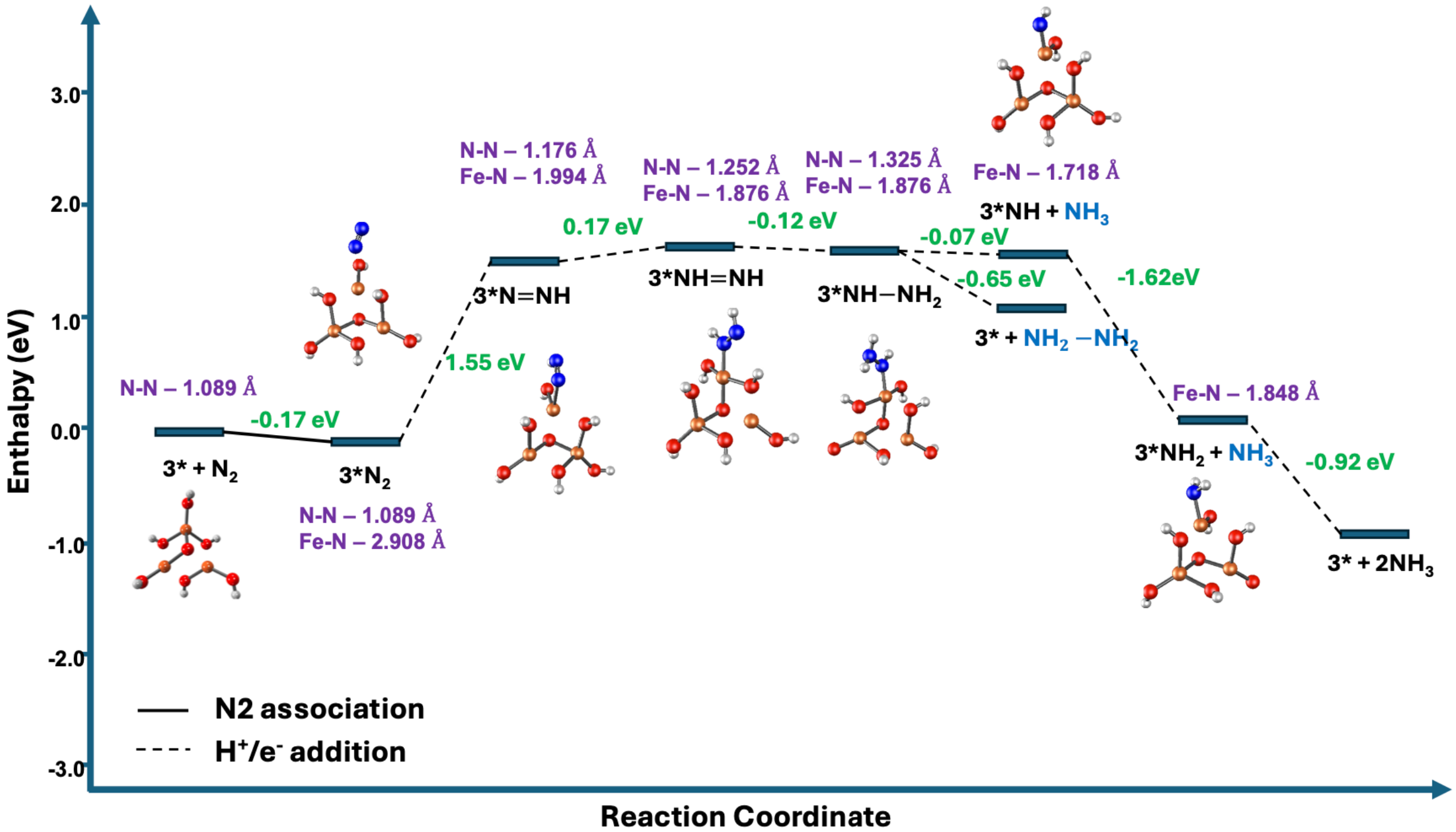

**Figure S16:** *Enthalpy change for the NRR mechanism on Fe₃O₄ after three water dissociations.* 3* refers to $Fe_3(OH)_3O_4(H)_3$.

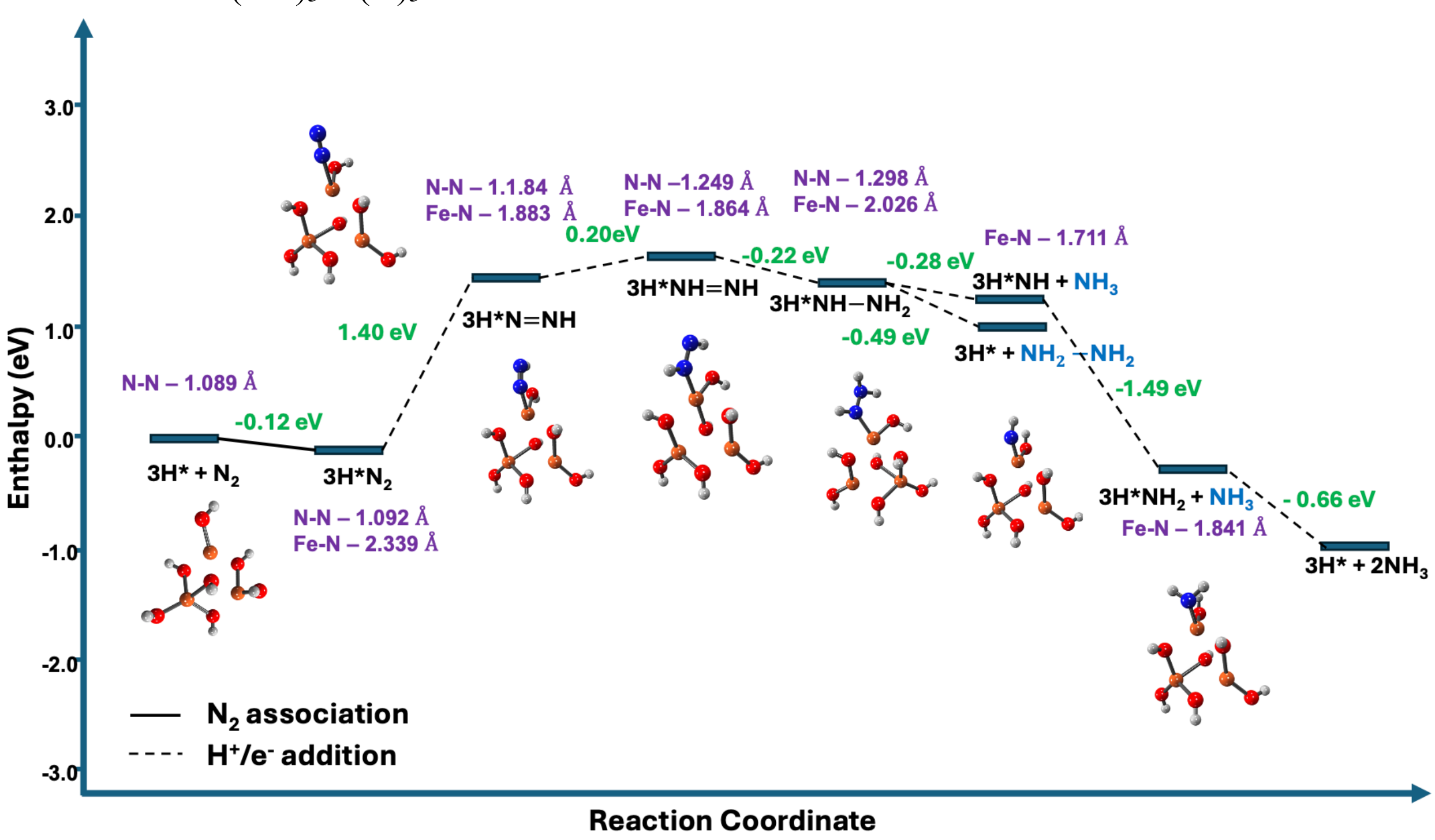

**Figure S17:** *Enthalpy change for the NRR mechanism on $Fe_3O_4$ after three water dissociations plus additional hydrogen/electron.* 3H* refers to $Fe_3(OH)_3O_4(H)_4$

**REFERENCES**

1. Fairley, N.; Fernandez, V.; Richard-Plouet, M.; Guillot-Deudon, C.; Walton, J.; Smith, E.; Flahaut, D.; Greiner, M.; Biesinger, M.; Tougaard, S.; Morgan, D.; Baltrusaitis, J. Systematic and collaborative approach to problem solving using X-ray photoelectron spectroscopy. *Appl Surf Sci Adv* **2021,** *5*.

2. Kendelewicz, T.; Kaya, S.; Newberg, J. T.; Bluhm, H.; Mulakaluri, N.; Moritz, W.; Scheffler, M.; Nilsson, A.; Pentcheva, R.; Brown, G. E., Jr. X-ray Photoemission and Density Functional Theory Study of the Interaction of Water Vapor with the Fe3O4(001) Surface at Near-Ambient Conditions. *The Journal of Physical Chemistry C* **2013,** *117* (6), 2719-2733.

3. Kraushofer, F.; Mirabella, F.; Xu, J.; Pavelec, J.; Balajka, J.; Müllner, M.; Resch, N.; Jakub, Z.; Hulva, J.; Meier, M.; Schmid, M.; Diebold, U.; Parkinson, G. S. Self-limited growth of an oxyhydroxide phase at the FeO(001) surface in liquid and ambient pressure water. *Journal of Chemical Physics* **2019,** *151* (15).

4. Zborowski, C.; Vanleenhove, A.; Hoflijk, I.; Vaesen, I.; Artyushkova, K.; Conard, T. High energy x-ray photoelectron spectroscopy Cr measurement of bulk gadolinium. *Surf Sci Spectra* **2023,** *30* (2).

5. Sanchez, M. B.; Huerta-Ruelas, J. A.; Cabrera-German, D.; Herrera-Gomez, A. Composition assessment of ferric oxide by accurate peak fitting of the Fe 2 photoemission spectrum. *Surf Interface Anal* **2017,** *49* (4), 253-260.

6. Yamashita, T.; Hayes, P. Analysis of XPS spectra of Fe and Fe ions in oxide materials. *Appl Surf Sci* **2008,** *254* (8), 2441-2449.

7. Grunze, M. The interaction of hydrazine with an Fe(111) surface. *Surface Science* **1979,** *81* (2), 603-625.

8. Tanaka, S.; Uyama, H.; Matsumoto, O. Synergistic Effects of Catalysts and Plasmas on the Synthesis of Ammonia and Hydrazine. *Plasma Chem Plasma P* **1994,** *14* (4), 491-504.

9. Alberas, D. J.; Kiss, J.; Liu, Z. M.; White, J. M. Surface-Chemistry of Hydrazine on Pt(111). *Surface Science* **1992,** *278* (1-2), 51-61.

10. Brumovsky, M.; Oborná, J.; Micic, V.; Malina, O.; Kaslík, J.; Tunega, D.; Kolos, M.; Hofmann, T.; Karlicky, F.; Filip, J. Iron Nitride Nanoparticles for Enhanced Reductive Dechlorination of Trichloroethylene. *Environ Sci Technol* **2022,** *56* (7), 4425-4436.

11. Degaga, G. D.; Trought, M.; Nemsak, S.; Crumlin, E. J.; Seel, M.; Pandey, R.; Perrine, K. A. Investigation of N(2) adsorption on Fe(3)O(4)(001) using ambient pressure X-ray photoelectron spectroscopy and density functional theory. *J Chem Phys* **2020,** *152* (5), 054717.